\documentclass[lettersize,journal]{IEEEtran}
\usepackage{amsmath,amsfonts}
\usepackage{algorithmic,hyperref}
\usepackage{algorithm}
\usepackage{array}
\usepackage{textcomp}
\usepackage{stfloats}
\usepackage{url}
\usepackage{verbatim}
\usepackage{graphicx}
\usepackage{cite}
\usepackage{color,soul}
\usepackage{subcaption}
\usepackage{amsfonts}
\usepackage{booktabs}
\usepackage{siunitx}
\usepackage{widetext}
\usepackage{epstopdf}
\DeclareMathOperator{\sech}{sech}

\begin{document}

\title{On  the  soliton  solutions  in  a self-gravitating  strongly  coupled electron-ion-dusty plasma}

\author{ Shatadru Chaudhuri, Shahin Nasrin and A. Roy Chowdhury
\IEEEcompsocitemizethanks{
     \IEEEcompsocthanksitem Shatadru Chaudhuri is with the Department of Physics, Shri Shikshayatan College, Kolkata - 700071, India\\
     Also at Department of Physics, Jadavpur University, Kolkata - 700032, India e-mail: (shatadruchaudhuri.phys.rs@jadavpuruniversity.in).
     \IEEEcompsocthanksitem Shahin Nasrin is with the Department of Physics, Shri Shikshayatan College, Kolkata - 700071, India
e-mail:(shahin.nas@gmail.com)
     \IEEEcompsocthanksitem A. Roy Chowdhury is with High Energy Physics Division, Department of Physics, Jadavpur University, Kolkata - 700032, India  e-mail: (arc.roy@gmail.com).}
\thanks{ Manuscript received \today; revised \today }}

\maketitle

\begin{abstract}
The effect of electrostatic strong-coupling of dust particles along with their self-gravitational force has been analyzed in a three component dusty plasma. The electrons and ions forming the charge neutral background where the electron distribution is assumed to be Maxwellian while the ion distribution is non-thermal. These days, one of the key topics in plasma physics is nonlinear waves in plasma.
Thus using the reductive perturbation technique to the set of hydrodynamic equation considered for an electron-ion-dusty (e-i-d) plasma, a coupled KdV equation is derived. The impact of strong coupling and self-gravitation on the solitary wave profiles, nonlinear coefficient and dispersive coefficient are studied both analytically and by numerical simulation.
\end{abstract}

\begin{IEEEkeywords}
Dusty Plasma, Gravitating Plasma, Strongly Coupled Plasma, Dust acoustic solitons
\end{IEEEkeywords}

\section{Introduction}\label{intro}
\IEEEPARstart{T}he studies of dusty plasma has become a crucial area of research due to the presence of dust in numerous natural environments, including radio-frequency discharges, near planets, and interstellar and circumstellar media. Due to its many applications to laboratory, space, and astrophysical plasma settings, such as asteroid zones, planatery rings, cometary tails, the interstellar medium, earth's surroundings, etc., interest in the field of dusty plasma physics has been expanding quickly\cite{fortov1997crystalline, fortov1996emission, goertz1989dusty}. In reality, the charge on the dust grain varies with time and space because of electron and ion currents that flow into and out of the grain in addition to other processes like electron photoemission and secondary emission. This causes fluctuations in the dust charge. When these dust grains collide with other charged particles, such electrons, they get charged and become submerged in the surrounding plasma. In the majority of astrophysical scenarios, a grain becomes charged due to the radiative and plasma backdrop it is submerged in. The grains' weights and sizes could very well be dependent on the same plasma environment, for instance. Larger and heavier grains could result from coulombic coagulation, but lighter and smaller grains could come from Maxwellian stress breaking apart the huge grains. These charged grains use electromagnetic (EM) forces to couple with the dynamics of the plasma. Low-frequency eigenmodes arise because the characteristic dust frequencies are significantly smaller than those of ions and electrons, due to the potentially very high effective charge and mass of these grains. Since each isolated dust grain has a mass that is typically between $10^{6}$ and $10^{12}$ times that of an ion\cite{mendis1979dust, misra2006dust}. A new ultra low-frequency domain for the development of various acoustic mode types in dusty plasmas — which are absent in typical electron-ion plasmas — is created by the presence of such rather big particles. One such significant mode is the dust-acoustic mode, an Eigenmode of the dust-electron-ion plasma in which the restoring force is provided by the pressures of inertialess electrons and ions and the inertia is provided by the charged dust grains. As a result, in many laboratory studies, the dust-acoustic waves (DAW) have been seen with the unaided eye on a kinetic level\cite{barkan1995laboratory, prabhakara1996trapping, thomas2001dust, fortov2003dust}. Dust acoustic waves, or DAWs, are the normal mode of a three-component dusty plasma made up of electrons, ions, and extremely huge micro-meter-sized charged dust grains. DAWs have a very low phase velocity (compared to the electron and ion thermal velocity). Star formation is one of the most significant of the many ancient issues that must be reconsidered in light of the many recent advances in our understanding of dusty plasma. The question of magneto-gravitational instability of stars is of great importance due to the competition between the gravitational and electromagnetic forces. Because of the increased mass of the grains, there is typically conflict between the gravitational and electromagnetic effects. Gravitational forces will prevail for very large grains, while electromagnetic and gravitational forces may compete for moderate-sized dusts. This equilibrium served as the foundation for the well-known Jeans instability, and serves as a standard for the disintegration of interstellar material. Interesting analyses of the modification of Jeans' criterion resulting from the interaction of nonlinearity, gravitational force, and electromagnetic effect have previously been conducted by a few writers\cite{abbasi2018effect, prajapati2013effect, 1976PASJ28371A}. Naturally, the tale of dusty plasma is fairly ancient, with several models being postulated for various kinds of investigation\cite{100292004RG000151, Merlino2006DustyPA}. The electrostatic energy of the mutual interaction is unusually high due to huge charges carried by the grains (usually of the order of thousand elementary charges for a micron-size particle). Therefore, compared to the electron-ion subsystem, it is far easier to achieve significant electrostatic coupling in the dust subsystem. Transitions from a disordered gaseous-like phase to a liquid-like phase and the creation of organized structures from dust particles, known as plasma crystals, are observed in complex plasmas. In 1959, Wuerker et al.\cite{wuerker1959electrodynamic} reported the first experimental observation of the ordered (quasicrystalline) structures of charged microparticles formed in a modified Paul's trap. It was Ikezi (1986)\cite{ikezi1986coulomb} who first hypothesized that dust subsystem crystallization may occur in a nonequilibrium gas discharge plasma.
\par Because of its many applications, the physics of strongly coupled plasmas, where the average potential energy per particle predominates over the average kinetic energy, is of enormous interest. When such high coupling is present, the presence of a huge charged dust component in a typical electron-ion plasma can have a significant impact on the dynamics of the system. Recently, both theoretical and experimental investigations of complicated dynamics have been conducted in such a plasma system. The coupling parameter ($\Gamma$) is one of the essential features of a system where many particle interactions occur. The parameter ‘$\Gamma$’ is characterized as the ratio of the potential energy of interaction between the neighboring particles to their kinetic energy. The coupling parameter is expressed as $\Gamma = \frac{q_{d}^{2}}{ak_{B}T_{d}}\exp[-\mathcal{K}]$, with $\mathcal{K}=a_{d}/\lambda_{D}$. The quantity `$\mathcal{K}$' is known as the screening parameter. $q_{_d}$ is the charge on each dust particle, $a_{_d}(\approx n_{d}^{-1/3})$ is the inter-dust  particle distance, $n_{d}$ and $T_{d}$ are the dust density and temperature, $\lambda_{D}$ is the plasma Debye length, $k_{B}$ the Boltzmann\index{Boltzmann} constant. The possibility of a classical Coulomb plasma with a population of dust grains transitioning to a strongly coupled phase was first noted by Ikezi\cite{ikezi1986coulomb}. It had been nearly twenty-five years ago. The reason for this is that with high charge and low temperature, the coupling parameters approach or even surpass 1. In fact, the Debye-Huckel repulsive potential determines the way the negatively charged micro-particles interact with one another. It has been shown that a strongly coupled plasma can behave like a liquid or a solid, with viscosity and elasticity playing equally significant roles. Dust grains are known to have a quasi-crystalline structure, and these characteristics are referred to as visco-elasticity. Therefore, there should be strong correlation effects influencing the low frequency collective modes in such a new phase of plasma.
\par Currently, there are two paths toward the development of a strongly linked plasma. Effective electrostatic force\cite{yaroshenko2009bohm,PhysRevE.86.066404,PhysRevE.89.043103,sultana2020dust} and visco-elastic effects\cite{banerjee2010viscoelastic,goswami2023revisiting,dharodi2024vortex} are the two types of forces. A number of these have applications in the study of charged particles in cryogenic traps\cite{gruber2005formation}, electrons trapped on the surface of liquid helium\cite{kalman1998strongly}, and other astrophysical systems\cite{merlino2004dusty}, including the ion-liquid inside white dwarf interiors, the crust of neutron stars, supernova cores, etc.
\par As the diameters of the dust grains increase, the gravitational effects become significant. Similar to other plasma particles, trapping by magnetic field lines predominates for grains that are too tiny. Conversely, excessively big grains do not deviate from their gravitationally constrained orbits due to electromagnetic forces. There could be a struggle between the electrostatic repulsion and gravitational self-attraction of the grains in plasmas such as protostellar clouds. Dusty plasmas are susceptible to macroscopic instabilities of the Jeans type\cite{PhysRevE.49.5599,doi:10.1098/rspa.2005.1594,MACE1998146,10.1063/1.56681}, when self-gravitational interaction resulting from the behavior of dust component is taken into account. In contrast to the electromagnetic force, the gravitational force is solely attracting, which accounts for the Jeans instability of a huge system.
\par El-Labany et al.\cite{el2003dust,el2004dust} recently obtained stationary modes, tiny amplitude DA solitons, and double layers while taking the dynamics of dust charge and trapped electrons/ions into account. It has already been demonstrated that the coexistence of rarefactive and compressive DA solitary structures and double layers is possible at the two-ion temperature. Nonetheless, these studies that disregard gravity are only applicable in the plasma phase, where the electrostatic force is significantly stronger than the gravitational pull.
\par In many astrophysical environments, especially in dense regions like star-forming molecular clouds and accretion disks, dusty plasmas are both self-gravitating and can exhibit strong coupling. The interplay of gravity, electrostatic forces, and plasma dynamics can lead to complex behaviors, including clustering, filamentation, and large-scale structure formation\cite{shukla2001survey}. Even in the laboratory, dusty plasmas are often studied under conditions where gravity can be minimized (such as in microgravity experiments aboard the International Space Station) to understand how dust particles behave in strongly coupled states\cite{usachev2014externally,petrov2013collective}.
\par While several studies have examined the impact of trapped electrons or ions, variations in dust temperatures, and charge fluctuations, none have examined the consequences of electrostatic strong coupling in conjunction with self-gravitation. Using an appropriate reductive perturbation technique, we have analyzed both the linear and nonlinear regimes of a self-gravitating quantum dusty plasma in the current communication. Two coupled modified KdV equations have been derived in the nonlinear section, while a new dispersion relation providing precise information about different propagation modes is generated in the linear regime.
\par The structure of our presentation is as follows: in Section \ref{formu}, we give the density distributions for the Maxwellian distributed electrons and non-thermal ions as well as the hydrodynamic equations for the strongly coupled as well as self-gravitating dust fluid. The goal of Section \ref{lnlanaly} is to derive the linear and nonlinear equations by analyzing different orders of perturbation terms that eventually makes us able to explain how large amplitude dust waves behave. 
The obtained results were discussed and explained precisely in Section\ref{resultsdiscuss} as well.

\section{Formulation}\label{formu}
Due to the fact that electrons and ions are collisionally attached to the grain surface, it is expected that a three component dusty plasma made up of electrons, ions, and charged dust grains has grains with the same charge and radius. The current study ignores the effects of photoemission, secondary emission, etc. It is important to note that this assumption does not fully account for the thermal grains. In the continuity and momentum exchange equations, sink and source terms are created by the charging and discharging of the grain. We think that the grains determine the gravitational potential, mostly because of the enormous mass differential between the grains and the plasma. The dynamics of this three component dusty plasma is described by the following model equation;

\begin{equation}\label{eqn1}
\partial_t n_{d}+\partial_x(n_{d}u_{d})=0
\end{equation}

\begin{equation}\label{eqn2}
  \partial_t u_{d}+u_{d}\partial_x u_{d}=\frac{Z_{d}e}{m_{d}}\partial_x \phi-\frac{1}{n_{d}}\partial_{x}P_{*} - \partial_{x}\psi
\end{equation}

\begin{equation}\label{eqn3}
 \partial^{2}_{xx} \phi=-\frac{e}{\varepsilon_{0}}(n_{i}-n_{e}-n_{d}Z_{d})
\end{equation}

\begin{equation}\label{eqn4}
 \partial^{2}_{xx} \psi=4\pi Gm_{d}n_{d} + 4\pi Gm_{i}n_{i}
\end{equation}

\begin{equation}\label{eqn5}
n_{e} = n_{e0}\exp\left[-\frac{e\phi}{K_{B}T_{e}}\right]
\end{equation}

\begin{equation}\label{eqn6}
n_{i} = n_{i0}\left[1+\beta\left(\frac{e\phi}{K_{B}T_{i}}\right)+\beta\left(\frac{e\phi}{K_{B}T_{i}}\right)^{2}\right]\exp\left(\frac{e\phi}{K_{B}T_{i}}\right)
\end{equation}

\begin{equation}\label{eqn7}
P_{*} = n_{d}K_{B}T_{*}
\end{equation}

Here, $\partial_t,\,\partial_x\text{ and }\partial^2_{xx}$ represent first order partial derivative with respect to time, space and second order partial derivative of space respectively. In the equation set (\ref{eqn1} - \ref{eqn7}) the symbols used denotes the following physical parameters viz. $n_{d}$ the number density of the dust particles, $u_{d}$ the velocity of the dust particles, $m_{d}$ the mass of the dust grains, $e$ the electronic charge, $Z_{d}$ the charge number associated with the dust grains. $n_{e}$ and $n_{i}$ are the number density of the electrons and ions which have temperatures $T_{e}$ and $T_{i}$ respectively. The electrons are assumed to be hot and hence follow the Maxwellian distribution while the ions are considered to be non-thermally distributed forming a charge neutral background for the dusty plasma system. It is to be mentioned in this context that, it is possible to have a plasma system with Maxwellian distributed electrons and non-thermal ions. In such a system, the electrons follow a Maxwell-Boltzmann distribution, which means their velocities and energies are distributed according to thermal equilibrium at a certain temperature. However, the ions in the plasma can deviate from this thermal equilibrium and exhibit non-thermal behaviour. Plasma waves or instabilities can interact with ions and transfer energy to them, producing non-thermal ion distributions. This can occur without significantly affecting the electron distribution. In plasmas with low collisionality, the ions may not have enough collisions to thermalize, while the lighter electrons can thermalize more quickly due to their higher collision rates.
\par The parameter $\beta = 4\alpha/(1 + 3\alpha)$, $\alpha$ being the non-thermal parameter, that indicates how far the behavior deviates from typical Maxwellian behavior. $\phi$ and $\psi$ are the electrostatic and gravitational potential and $P_{*}$ known as the electrostatic pressure, arising due to the strong coupling between the dust particles whose mathematical expression comes from an equation of state defined by Gozadinos et.al.\cite{gozadinos2003fluid} as;

\begin{equation}\label{eqn8}
P_{*} \simeq \frac{N_{nn}}{3}\Gamma K_{B}T_{d}n_{d}(1 + \mathcal{K})\exp[-\mathcal{K}]
\end{equation}

Later a parameter defined by Cousens et.al.\cite{PhysRevE.86.066404} in this context as the `effective electrostatic temperature', $T_{*}$ which is generally some orders of magnitude higher than the kinetic temperature of dust ($T_{d}$). According to them the aforementioned can be mathematically expressed as,

\begin{equation}\label{eqn9}
K_{B}T_{*} = \frac{N_{nn}Z_{d}^{2}e^{2}}{12\pi\varepsilon_{0}}\sqrt[3]{n_{d}}(1+\mathcal{K})\exp[-\mathcal{K}]
\end{equation}

where, the parameter ``$N_{nn}$" refers to the number of closest neighbours that establishes the structure of the dusty plasma.

\par Before proceeding to the linear and nonlinear analysis we need to make the dynamical variables of the equation set (\ref{eqn1}-\ref{eqn7}) dimensionless. This can be achieved by normalization of the variables by some plasma parameters which are as follows;\\
The dust number density is normalized as $n \rightarrow n_{d}/n_{d0}$, the electron and the ion density is normalized as $n_{l} \rightarrow n_{l}/n_{l0}$, where `$l$' stands for the type of species (i.e. $l$ = $e$ or $i$ for electron or ion). The dust velocity is normalized as $u \rightarrow u_{d}/c_{d}$. The space and the time are normalized as $x \rightarrow x/\lambda_{D}$ and $t \rightarrow \omega_{pd}t$ respectively. The electrostatic and the gravitation potentials are normalized as $\phi \rightarrow \phi/\left(\frac{K_{B}T_{e}}{e}\right)$ and $\psi \rightarrow \psi/c_{d}^{2}$. Last but not the least the effective temperature is normalized as $T_{*}\rightarrow T_{*}/T_{0}$.
The expressions of the normalizing constants are given by the Table-\ref{tab1}. Thus the normalized governing equations are as follows;
\begin{equation}\label{eqn10}
  \partial_t n_{d}+\partial_x(n_{d}u_{d})=0
\end{equation}

\begin{equation}\label{eqn11}
  \partial_t u + u\partial_x u=\partial_x \phi-n^{-1}\partial_{x}(nd)-\partial_{x}\psi
\end{equation}

\begin{equation}\label{eqn12}
 \partial^{2}_{xx} \phi=(n + \mu_{e}n_{e} - \mu_{i}n_{i})
\end{equation}

\begin{equation}\label{eqn13}
 \partial^{2}_{xx}\psi=\gamma_{jd}^{2}n + \gamma_{ji}^{2}n_{i}
\end{equation}

\begin{equation}\label{eqn14}
n_{e} = \exp\left[\sigma_{e}\phi\right]
\end{equation}

\begin{equation}\label{eqn15}
n_{i} = \left[1+ \Lambda_{i}\phi + \Lambda_{i}^{2}\phi_{2}\right]\exp\left(-\sigma_{i}\phi\right)
\end{equation}

`$\mu_{e}(=\frac{n_{e0}}{Z_{d}n_{d0}})$' and `$\mu_{i}(=\frac{n_{i0}}{Z_{d}n_{d0}})$' in the equation(\ref{eqn12}) are the equilibrium density ratios for electron and ion respectively. In equation(\ref{eqn13}), $\gamma_{jd}^{2}=\left(\omega_{jd}/\omega_{pd}\right)^{2}$ and $\gamma_{ji}^{2}=\left(\omega_{ji}/\omega_{pd}\right)^{2}$ are the normalized Jeans frequency for dust and ion respectively, whereas $\omega_{jd}^{2}=4\pi Gm_{d}n_{d}$ and $\omega_{ji}^{2}=4\pi Gm_{i}n_{i}$ are the Jeans frequency for dust and ion respectively. $\sigma_{e}(=T_{0}/T_{e}Z_{d})$ in equation(\ref{eqn14}) is the normalized dust-electron temperature ratio. Similarly, in equation(\ref{eqn15}) the parameter $\sigma_{i}(=T_{0}/T_{i}Z_{d})$ is the normalized dust-ion temperature ratio. While, another parameter $\Lambda_{i}$ arising as coefficient of $\phi$ (or $\phi^{2}$) corresponds to a simple product of constants, i.e. $\Lambda_{i}=\beta\sigma_{i}$. One important thing is worth mentioning in this case is that, after normalization of the equation(\ref{eqn2}) an important dynamical variable gets associated to equation(\ref{eqn11}) which is `$d$'. This dynamical variable comes as a result of the normalization of the effective temperature ($T_{*}$). So, $d=T_{*}/T_{0}$. Since, $T_{*}$ (as per \ref{eqn9}) is a function of ($n,\phi$) hence, $d$ is a dynamically varying quantity.
\par Before proceeding to the next section one very important thing to mention is that, following the analysis of Cousens et.al.\cite{PhysRevE.86.066404} we use;
\begin{equation}\label{eqn16}
\begin{split}
   & \left. \begin{array}{l}
   d_{1} = d_{11}n_{1} + d_{12}\phi_{1} \\
   d_{2} = d_{21}n_{2} + d_{22}\phi_{2} + d_{23}n_{1}^{2} + d_{24}n_{1}\phi_{1} + d_{25}\phi_{1}^{2} \\
   \end{array} \right\}
   \end{split}
\end{equation}

\begin{table}
\caption{\label{tab1}}
\centering
\resizebox{\linewidth}{!}{
\begin{tabular}{ccc} \toprule
    Sl.No. & Plasma Parameters & Mathematical Expressions \\ 
    \midrule
    1  & Temperature & $T_{0} = \sqrt{\frac{Z_{d}^{2}n_{d0}T_{i}T_{e}}{n_{e0}T_{i} + n_{i0}T_{e}}}$ \\
    2  & Debye length  & $\lambda_{D} = \sqrt{\frac{\varepsilon_{0}K_{B}T_{0}}{n_{d0}Z_{d}^{2}e^{2}}}$ \\
    3  & Dust frequency  & $\omega_{pd}=\sqrt{\frac{n_{d0}Z_{d}^{2}e^{2}}{\varepsilon_{0}m_{d}}}$ \\
    4  & Dust acoustic speed  & $c_{d}=\sqrt{\frac{K_{B}T_{0}}{m_{d}}}$ \\
    5  & Equilibrium electron number density  & $n_{e0}$ \\ 
    6  & Equilibrium ion number density  & $n_{i0}$ \\
    \midrule
   \bottomrule
\end{tabular}}
\end{table}
\section{Linear and Non-linear Analysis}\label{lnlanaly}
In order to study the wave-wave and wave-particle interactions in this three component dusty plasma we use the reductive perturbation technique and apply it to the equation set (\ref{eqn10} - \ref{eqn15}). In this present presentation, we are focusing our analysis on the study of wave excitation occurring in this electron-ion-dusty plasma for the long wavelength approximation. Thus, we proceed by implementing the following stretching of the coordinates, viz.
\begin{equation}\label{eqn17}
\begin{split}
   & \left. \begin{array}{c}
   \xi = \epsilon^{1/2}(x - v_{p}t) \\
   \tau = \epsilon^{3/2}t \\
   \end{array} \right\}
   \end{split}
\end{equation}
Next, we consider the following perturbation of the dynamical variables;
\begin{equation}\label{eqn18}
\begin{split}
   & \left. \begin{array}{l}
   n = 1 + \epsilon n_{1} + \epsilon^{2}n_{2} + \epsilon^{3}n_{3} + \cdots\\
   u = \epsilon u_{1} + \epsilon^{2}u_{2} + \epsilon^{3}u_{3} + \cdots\\
   \phi = \epsilon \phi_{1} + \epsilon^{2}\phi_{2} + \epsilon^{3}\phi_{3} + \cdots\\
   \psi = \epsilon \psi_{1} + \epsilon^{2}\psi_{2} + \epsilon^{3}\psi_{3} + \cdots\\
   d = d_{0} + \epsilon d_{1} + \epsilon^{2}d_{2} + \epsilon^{3}d_{3} + \cdots
   \end{array} \right\}
   \end{split}
\end{equation}
For the zeroth order of the $\epsilon$ from the equation(\ref{eqn12}) we obtain the quasi-neutrality condition which is,
\begin{equation}\label{eqn19}
n_{i0} = n_{e0} + Z_{d}n_{d0}
\end{equation}
Using the equation set (\ref{eqn10} - \ref{eqn17}) and equating the first order of $\epsilon$ i.e. for $\epsilon^{3/2}$ we get from equation(\ref{eqn10}) and (\ref{eqn11}) the set of first order equations, which are as follows;
\begin{equation}\label{eqn20}
n_{1} = \frac{1}{v_{p}}u_{1}
\end{equation}
\begin{equation}\label{eqn21}
u_{1} = \frac{v_{p}}{v_{p}^{2} + (d_{0}+d_{11})}\left[(d_{12}-1)\phi_{1} - \psi_{1}\right]
\end{equation}
Likewise equating the lowest order of $\epsilon$ (which is $\epsilon^{1}$) for equations (\ref{eqn12}) and (\ref{eqn13}) and incorporating the expressions (\ref{eqn14}) and (\ref{eqn15}) in them we get;
\begin{equation}\label{eqn22}
n_{1} = \left(\mu_{1}\theta_{1} - \mu_{e}\sigma_{e}\right)\phi_{1}
\end{equation}
\begin{equation}\label{eqn23}
\gamma_{jd}^{2}n_{1} + \gamma_{ji}^{2}\theta_{1}\phi_{1} = 0
\end{equation}
where, in the above two equations $\theta_{1} = -\sigma_{i} + \Lambda_{1}$
Using the equations (\ref{eqn20} - \ref{eqn23}) we get a pair of equation which is as follows;
\begin{equation}\label{eqn24}
E_{1}\phi_{1} - F_{1}\psi_{1} = 0
\end{equation}
\begin{equation}\label{eqn25}
E_{2}\phi_{1} - F_{2}\psi_{1} = 0
\end{equation}
where the expressions of $E_{1}$, $E_{2}$, $F_{1}$ and $F_{2}$ are given in the appendix portion.
\par The equation (\ref{eqn24}) and (\ref{eqn25}) demands that they can only be valid when,
\begin{equation}\label{eqn26}
\left|\begin{array}{cc}
E_{1} & -F_{1}\\
E_{2} & -F_{2}\\
\end{array}\right| = 0
\end{equation}
Solving equation(\ref{eqn26}) we get,
\begin{equation}\label{eqn27}
v_{p}^{2} = \frac{(1-\gamma_{jd}^{2})(d_{12}-1)}{\gamma_{ji}^{2} - (\mu_{i}\theta_{1}-\mu_{e}\sigma_{e})}+(d_{0} + d_{11})
\end{equation}
The equation(\ref{eqn27}) gives the phase velocity of the propagating wave.

\par Now we proceed to the higher order of $\epsilon$ and from the equation set (\ref{eqn10} - \ref{eqn13}) we get the following set of equations,
\begin{equation}\label{eqn28}
  \frac{\partial n_{1}}{\partial\tau} - v_{p}\frac{\partial n_{2}}{\partial\xi} + \frac{\partial u_{2}}{\partial\xi} + \frac{\partial}{\partial\xi}(n_{1}u_{1}) = 0
\end{equation}
\begin{equation}\label{eqn29}
\begin{split}
  &\frac{\partial u_{1}}{\partial\tau} - v_{p}\frac{\partial u_{2}}{\partial\xi} + u_{1}\frac{\partial u_{1}}{\partial\xi} = \frac{\partial\phi_{2}}{\partial\xi} - \frac{\partial d_{2}}{\partial\xi} + n_{1}\frac{\partial n_{1}}{\partial\xi} + d_{0}n_{1}\frac{\partial n_{1}}{\partial\xi} - \\
  &d_{0}\frac{\partial n_{2}}{\partial\xi} - \frac{\partial}{\partial\xi}(n_{1}d_{1}) - \frac{\partial\psi_{2}}{\partial\xi}
  \end{split}
\end{equation}
\begin{equation}\label{eqn30}
  \frac{\partial^{2}\phi_{1}}{\partial\xi^{2}} = n_{2} + \mu_{e}\sigma_{e}\phi_{2} + \frac{\sigma_{e}^{2}}{2}\phi_{1}^{2} - \mu-{i}\theta_{1}\phi_{2} - \mu_{i}\theta_{2}\phi_{1}^{2}
\end{equation}
\begin{equation}\label{eqn31}
  \frac{\partial^{2}\psi_{1}}{\partial\xi^{2}} = \gamma_{jd}^{2}n_{2} + \gamma_{ji}^{2}\theta_{1}\phi_{2} + \gamma_{Ji}^{2}\theta_{2}\phi_{1}^{2}
\end{equation}
Using equation set (\ref{eqn20} - \ref{eqn25}) and (\ref{eqn27} - \ref{eqn31}) we get the following pair of equation,
\begin{equation}\label{eqn32}
\begin{split}
  &E_{1}\frac{\partial\phi_{2}}{\partial\xi} - F_{1}\frac{\partial\psi_{2}}{\partial\xi} = F_{1}\left[\frac{\partial^{3}\phi_{1}}{\partial\xi^{3}} - d_{23}\frac{\partial n_{1}^{2}}{\partial\xi} - d_{24}\frac{\partial}{\partial\xi}(n_{1}\phi_{1})\right.\\
  &\left.- d_{25}\frac{\partial}{\partial\xi}(\phi_{1}^{2}) + d_{11}n_{1}\frac{\partial n_{1}}{\partial\xi} + d_{12}n_{1}\frac{\partial\phi_{1}}{\partial\xi} + d_{0}n_{1}\frac{\partial n_{1}}{\partial\xi}\right. \\
  & - \left.\frac{\partial}{\partial\xi}(d_{11}n_{1}^{2} + d_{12}n_{1}\phi_{1}) +v_{p}\frac{\partial n_{1}}{\partial\tau}- v_{p}\frac{\partial}{\partial\xi}(n_{1}u_{1})\right. \\
  &\left.- \frac{\partial u_{1}}{\partial\tau} - u\frac{\partial u_{1}}{\partial\xi}\left(\frac{\sigma_{e}^{2}}{2}- \mu_{1}\theta_{2}\right)\frac{\partial\phi_{1}^{2}}{\partial\xi}\right]
  \end{split}
\end{equation}
\begin{equation}\label{eqn33}
\begin{split}
&F_{2}\frac{\partial\phi_{2}}{\partial\xi} - F_{2}\frac{\partial\psi_{2}}{\partial\xi} =\\\ &-\frac{\partial^{3}\psi_{1}}{\partial\xi^{3}} - E_{2}\left\{\frac{\partial}{\partial\xi}\left(d_{23}n_{1}^{2} + d_{24}n_{1}\phi_{1} + d_{25}\phi_{1}^{2}\right) + \right.\\
&\left. d_{11}n_{1}\frac{\partial n_{1}}{\partial\xi} + d_{12}n_{1}\frac{\partial\phi_{1}}{\partial\xi} + d_{0}n_{1}\frac{\partial n_{1}}{\partial\xi}-d_{11}\frac{\partial n_{1}^{2}}{\partial\xi} - \right.\\
&d_{12}\frac{\partial}{\partial\xi}(n_{1}\phi_{1})\left.v_{p}\frac{\partial n_{1}}{\partial\tau} + v_{p}\frac{\partial}{\partial\xi}(n_{1}u_{1}) + \frac{\partial u_{1}}{\partial\tau} + u_{1}\frac{\partial u_{1}}{\partial\xi}\right\} + \\
&2\gamma_{ji}^{2}\theta_{2}\phi_{1}\frac{\partial\phi_{1}}{\partial\xi}
\end{split}
\end{equation}
The equation (\ref{eqn32}) and (\ref{eqn33}) can be regarded as two linear homogeneous equations of `$\left(\frac{\partial\phi_{2}}{\partial\xi}\right)$' and `$\left(\frac{\partial\psi_{2}}{\partial\xi}\right)$'. Infact, we see that for non-zero solutions and by the virtue of equation (\ref{eqn26}) the right-hand-side of the equations (\ref{eqn32}) and (\ref{eqn33}) must vanish, which leads to
\section*{}
\begin{widetext}
\begin{equation}\label{eqn34}
\frac{\partial^{3}\phi_{1}}{\partial\xi^{3}} + A_{1}\phi_{1}\frac{\partial\phi_{1}}{\partial\xi} + B_{1}\frac{\partial\phi_{1}}{\partial\tau} + C_{1}\frac{\partial}{\partial\xi}(\phi_{1}\psi_{1}) + D_{1}\psi_{1}\frac{\partial\psi}{\partial\xi} + G_{1}\frac{\partial\psi_{1}}{\partial\tau}=0
\end{equation}
\begin{equation}\label{eqn35}
\frac{\partial^{3}\psi_{1}}{\partial\xi^{3}} + \bar{A}_{1}\phi_{1}\frac{\partial\phi_{1}}{\partial\xi} + \bar{B}_{1}\frac{\partial\phi_{1}}{\partial\tau} + \bar{C}_{1}\frac{\partial}{\partial\xi}(\phi_{1}\psi_{1}) + \bar{D}_{1}\psi_{1}\frac{\partial\psi}{\partial\xi} + \bar{G}_{1}\frac{\partial\psi_{1}}{\partial\tau}=0
\end{equation}
\end{widetext}
The equations (\ref{eqn34}) and (\ref{eqn35}) are two coupled KdV-equations where, the coefficients $A_{1}$, $B_{1}$ etc. are functions of plasma parameters and their expressions are mentioned in the appendix part of this paper.
\section{Solution of the Coupled KdV equation}\label{solckdv}
In order to seek the solution of nonlinear equations (\ref{eqn34}) and (\ref{eqn35}) one can consider $\psi_{1}$ and $\phi_{1}$ of the form
\begin{equation}\label{eqn36}
\left.\begin{array}{c}
\phi_{1} = a\sech^{2}(\xi - U\tau)\\
\psi_{1} = b\sech^{2}(\xi - U\tau)
\end{array}\right\}
\end{equation}
where, `$a$' and `$b$' are the amplitude of the soliton corresponding to the electrostatic and gravitational potential respectively. Here, `$U$' is the solitary wave's velocity and can be expressed as
\begin{equation}\label{eqn37}
\left.\begin{array}{c}
U = \frac{4}{B_{1}}\\
b = -\left(\bar{B}_{1}/B_{1}\right)a
\end{array}\right\}
\end{equation}
The expression of `$a$' reads as,
\begin{equation}\label{eqn38}
a = \frac{12\bar{B}_{1}}{B_{1}\left[\bar{A}_{1}-2C_{1}(\bar{B}_{1}/B_{1})-\bar{D}_{1}\bar{B}_{1}^{2}/B_{1}^{2}\right]}
\end{equation}
It is to be noted that the coefficients of the coupled KdV-equation are connected by the following equation,
\begin{equation}\label{eqn39}
\begin{split}
&\bar{B}_{1}A_{1} - B_{1}\bar{A_{1}} - 2\frac{\bar{B}_{1}}{B_{1}}\left[C_{1}\bar{B}_{1}-\bar{C}_{1}B_{1}\right]-\\
&\left(\frac{\bar{B}_{1}}{B_{1}}\right)^{2}(C_{1}\bar{B}_{1}-\bar{C}_{1}B_{1})=0
\end{split}
\end{equation}
The equation(\ref{eqn39}) can be regarded as a condition or a kind of restriction to the plasma parameter values defining the concerned plasma system. At this point one should note that the equation set (\ref{eqn37}) yields, 
\begin{equation}\label{eqn40}
b = -\frac{1}{4}\bar{B}_{1}Ua
\end{equation}
The amplitude of the  $\phi_{1}$'s soliton falls as $1/U$ when the $\psi_{1}$ soliton's amplitude increases with the wave's velocity, which is a typical soliton property. Conversely, the condition (\ref{eqn39}) suggests that specific circumstances must be met by the relevant plasma characteristics in order for such coupled solitons to arise. Thus, the basic characteristics of nonlinear waves in a plasma may vary depending on the influence of gravitation-like forces. Examining the function of Tsytovich's\cite{101071P98007} new kind of force, which will impact soliton formation differently, will undoubtedly be more fascinating.
\section{Results and Discussion}\label{resultsdiscuss}
\begin{figure}
    \centering
    \begin{subfigure}[b]{0.45\textwidth}
        \centering
        \includegraphics[width=\textwidth]{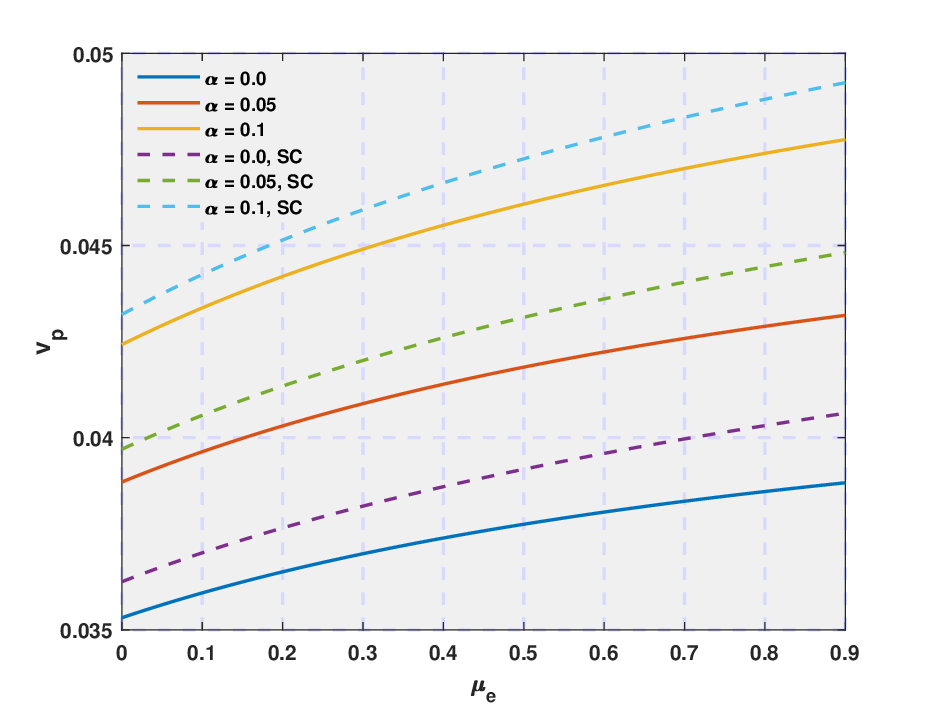}
        \caption{The variation of the phase velocity with the non-thermal parameter ($\alpha$). Here, $n_{d0} \sim 10^{12}m^{-3}$, $Z_{d} \sim 10^{3}$ and $\gamma_{jd} = 1.5$}
        \label{phasevelocity1}
    \end{subfigure}
    \hfill
    \begin{subfigure}[b]{0.45\textwidth}
        \centering
        \includegraphics[width=\textwidth]{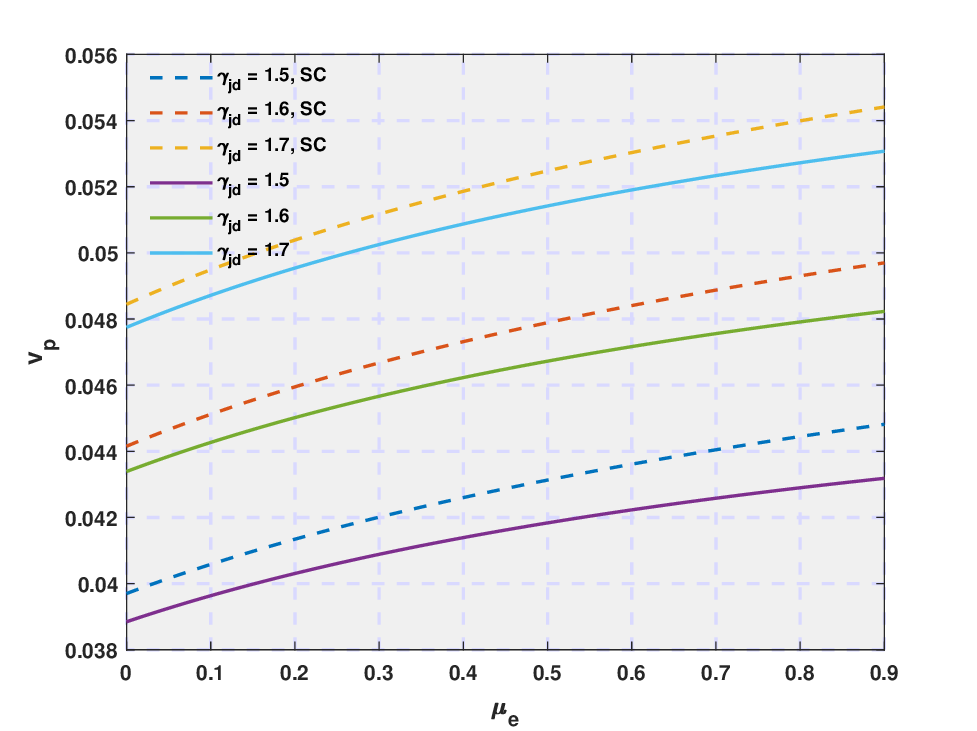}
        \caption{The variation of the phase velocity with the normalized dust Jeans frequency ($\gamma_{jd}$). Here, $\alpha = 0.2$, $Z_{d} \sim 10^{3}$, and $n_{d0} \sim 10^{12}$.}
        \label{phasevelocity2}
    \end{subfigure}
    \caption{The figure shows the variation of phase velocity ($v_{p}$) as a function of the electron-dust density ratio ($\mu_{e}$).}
    \label{phasevelocity}
\end{figure}

In section (\ref{formu}, \ref{lnlanaly}, \ref{solckdv}) we have executed some mathematical analysis and initially deriving the expression of phase velocity of the wave we ultimately came up with the derivation of a pair of coupled KdV equation. It is important to note at the beginning that while carrying out the whole analysis we have used a convection during the plot of graphs which is as follows :
\begin{enumerate}
\item The solid lines represents the graph of the plasma variable in absence of strong coupling.
\item The dashed lines represent the graph of the plasma variable in presence of strong coupling.
\end{enumerate}
To start with we focus on the expression of the phase velocity (\ref{eqn27}), which clearly shows that its nature of variation clearly depends upon the plasma parameters. Thus, to observe the nature of variation of the phase velocity ($v_{p}$) we have plotted $v_{p}$ as function of the equilibrium electron-dust density ratio ($\mu_{e}$) and the dust charge number ($Z_{d}$) represented by the figure (\ref{phasevelocity1} \& \ref{phasevelocity2}).
Both the plots of figure (\ref{phasevelocity}) clear the fact the with the increase of the initial number density of the electrons ($n_{e0}$) the phase velocity ($v_{p}$) increases. The figure (\ref{phasevelocity1}) shows that the magnitude of $v_{p}$, both in presence and absence of strong coupling, increases as $\alpha$ increases. This implies that the more the number of ions deviating from thermal equilibrium, the more the phase velocity is acquired by the wave. In next figure (i.e figure (\ref{phasevelocity2})), the effect of gravitational field on the phase velocity is also quite clear. In fact, with the increase in the value of the normalized Jeans frequency ($\gamma_{jd}$) the magnitude of phase velocity increases. This seems that the gravitational field making the motion of the dust particles more and more rigorous. Overall, we see that both the presence or absence of strong coupling only alters the numerical value of $v_{p}$ but the effect of nonthermal parameter and the gravitational field has a significant effect on the variation of the phase velocity.
\begin{figure}
    \centering
    \begin{subfigure}[b]{0.45\textwidth}
        \centering
        \includegraphics[width=\textwidth]{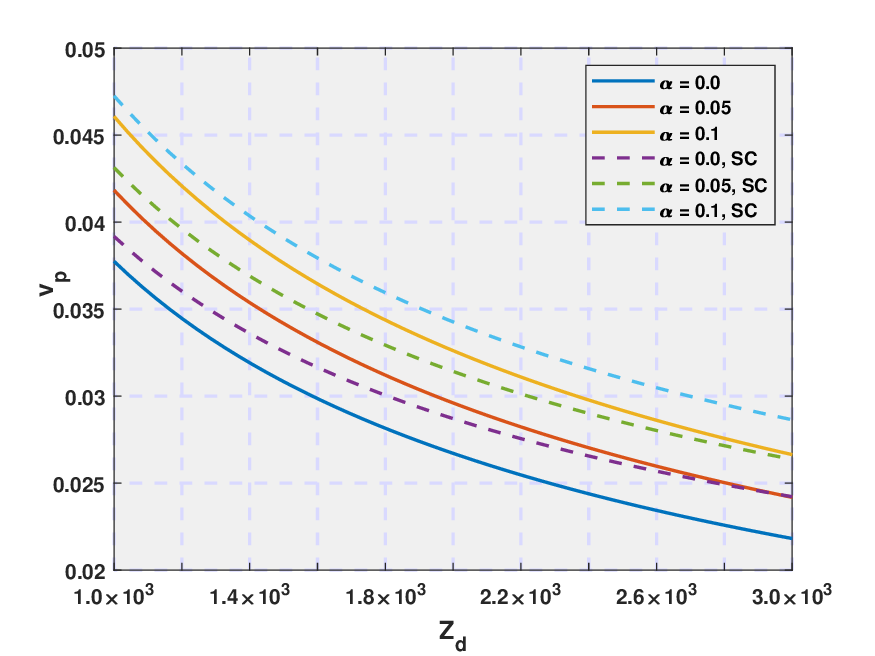}
        \caption{The variation of the phase velocity with the non-thermal parameter ($\alpha$). Here we set, $n_{d0} \sim 10^{12}m^{-3}$, $Z_{d} \sim 10^{3}$ and $\gamma_{jd} = 1.5$}
        \label{phasevelocity3}
    \end{subfigure}
    \hfill
    \begin{subfigure}[b]{0.45\textwidth}
        \centering
        \includegraphics[width=\textwidth]{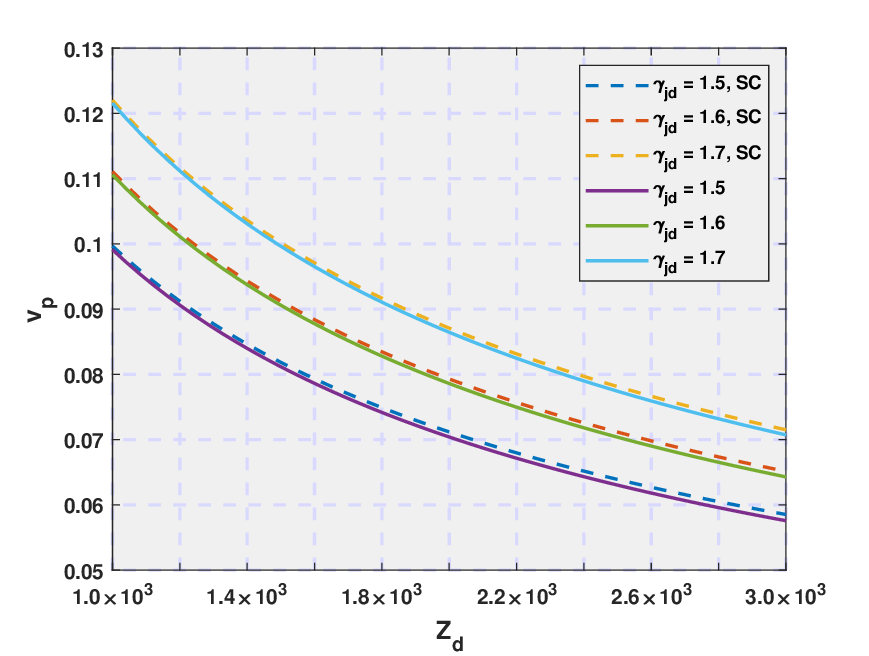}
        \caption{The variation of the phase velocity with the normalized dust Jeans frequency ($\gamma_{jd}$). The other parameters being constant and have the values such as, $\alpha = 0.2$, $Z_{d} \sim 10^{3}$, and $n_{d0} \sim 10^{12}$.}
        \label{phasevelocity4}
    \end{subfigure}
    \caption{The figure shows the variation of phase velocity ($v_{p}$) as a function of the dust charge number ($Z_{d}$).}
    \label{phasevelocity11}
\end{figure}
Furthermore, we have also observed that $v_{p}$ even has a dependence on the dust charge number ($Z_{d}$). The subplots of the figure (\ref{phasevelocity11}) clearly describes the dependence of $v_{p}$ on $Z_{d}$. The figure (\ref{phasevelocity3}) shows the variation of the phase velocity with $\alpha$ and $Z_{d}$. It's interesting to note that, $v_{p}$ increases with $\mu_{e}$ while decreases with $Z_{d}$ and at the end becomes constant at certain value. The figure depicts that though the phase velocity has a decreasing nature with respect to $Z_{d}$ but as $\alpha$ increases both the maxima and minima points of $v_{p}$ gets increased both in absence and presence of strong coupling.  This implies the same fact as before, that the more the number of fast moving ions, the more the phase velocity acquired by the wave.  Next when we switch to the figure (\ref{phasevelocity4}) we see that though the nature of variation of $v_{p}$ is the same with $Z_{d}$ but with change of $\gamma_{jd}$ there is no such variation occurring due to the presence or absence of strong coupling effect. Of course, the increases of normalized dust Jeans frequency ($\gamma_{jd}$) the phase velocity increases, which indicates that the interactions of the gravitation field with the dust particles causes the wave energy to get increased.
\begin{figure*}[h]
    \centering
    \includegraphics[width=1.0\textwidth,]{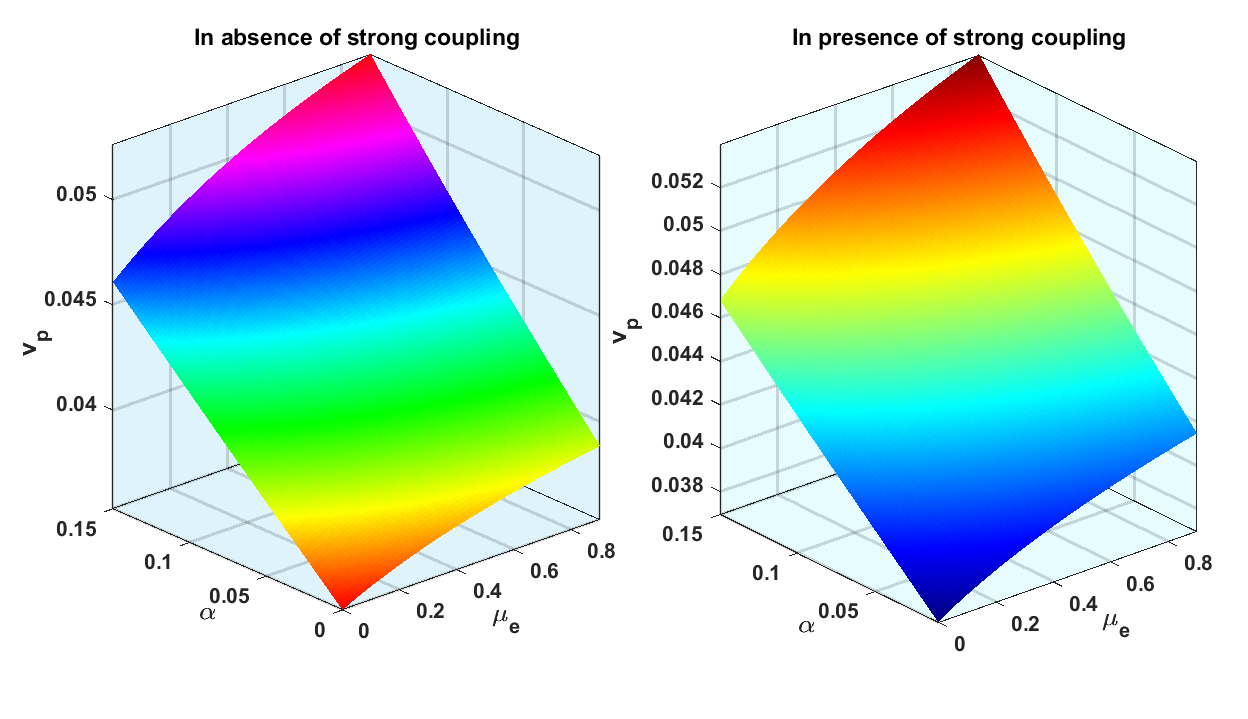}
    \caption{A 3D plot of phase velocity ($v_{p}$) showing the its dependence on $\mu_{e}$ and $\alpha$ as well. The values of $n_{d0}$ and $Z_{d}$ are kept at $10^{12}$ and $10^{3}$ respectively.}
    \label{subplot1}
\end{figure*}

The figure \ref{subplot1} (left) shows how the phase velocity vary with respect to $\alpha$ and $\mu_{e}$ simultaneously in absence of strong coupling. It is interesting to note that though $v_{p}$ rises much rapidly and linearly with $\alpha$ but its variation with $\mu_{e}$ is not linear.  This indicates, that the non-thermality of the ions have comparatively more impact on the wave-wave and wave-particle interaction of the concerned plasma system in compare to the number density of the electrons. In the figure (\ref{subplot1}) (right) the same is observed but in presence of strong coupling which depicts that its presence just increases the magnitude of $v_{p}$ implying that the wave energy gets increased due to the strong coupling effect.

\begin{figure}
    \centering
    \begin{subfigure}[b]{0.45\textwidth}
        \centering
        \includegraphics[width=\textwidth]{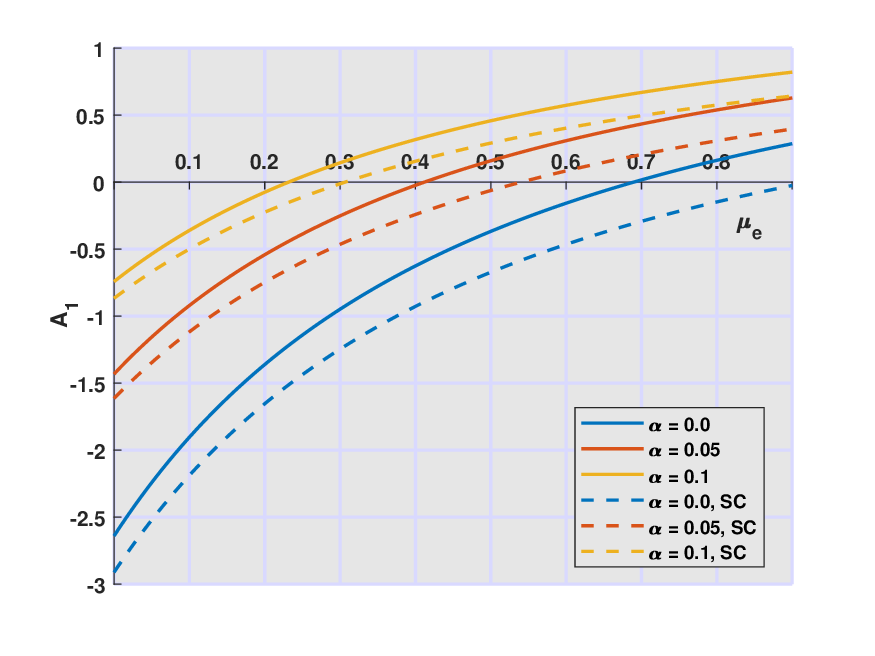}
        \caption{The variation of $A_{1}$ with the non-thermal parameter ($\alpha$). Here we set, $n_{d0} \sim 10^{12}m^{-3}$, $Z_{d} \sim 10^{3}$ and $\gamma_{jd} = 1.5$}
        \label{A1vsMue3}
    \end{subfigure}
    \hfill
    \begin{subfigure}[b]{0.45\textwidth}
        \centering
        \includegraphics[width=\textwidth]{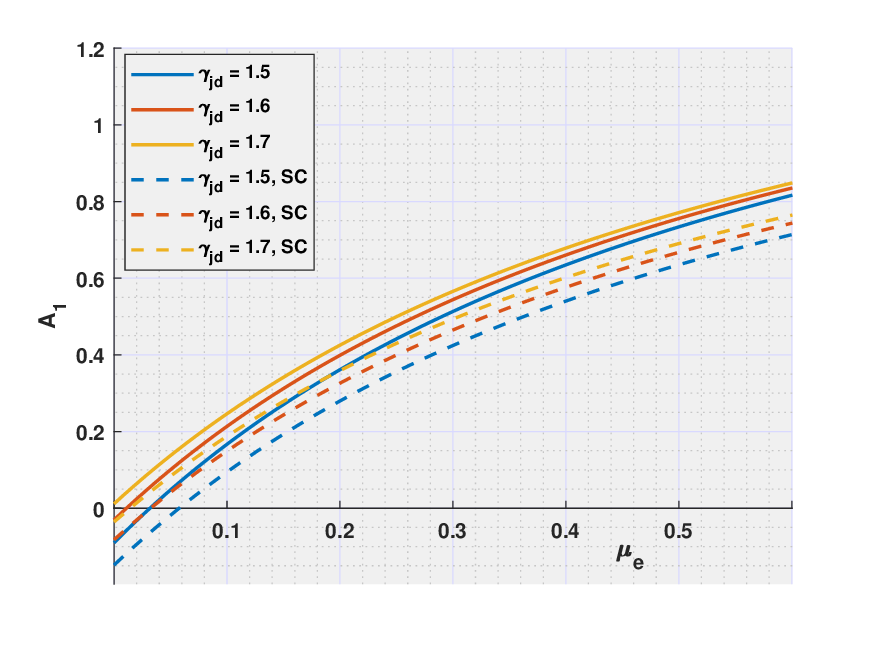}
        \caption{The variation of the $A_{1}$ with the normalized Jeans frequency for dust ($\gamma_{jd}$). The other parameters being constant and have the values such as, $\alpha = 0.2$, $Z_{d} \sim 10^{3}$, and $n_{d0} \sim 10^{12}$.}
        \label{A1vsMue4}
    \end{subfigure}
    \caption{The figure shows the variation of nonlinear coefficient ($A_{1}$) as a function of the electro-dust density ratio ($\mu_{e0}$). The coefficient governs the rate at which the wave steepens. Nonlinearity causes the crest of the wave to move faster than the trough, leading to wave distortion and steepening.}
    \label{A1vsMue}
\end{figure}

\par
\begin{figure}
    \centering
    \begin{subfigure}[b]{0.45\textwidth}
        \centering
        \includegraphics[width=\textwidth]{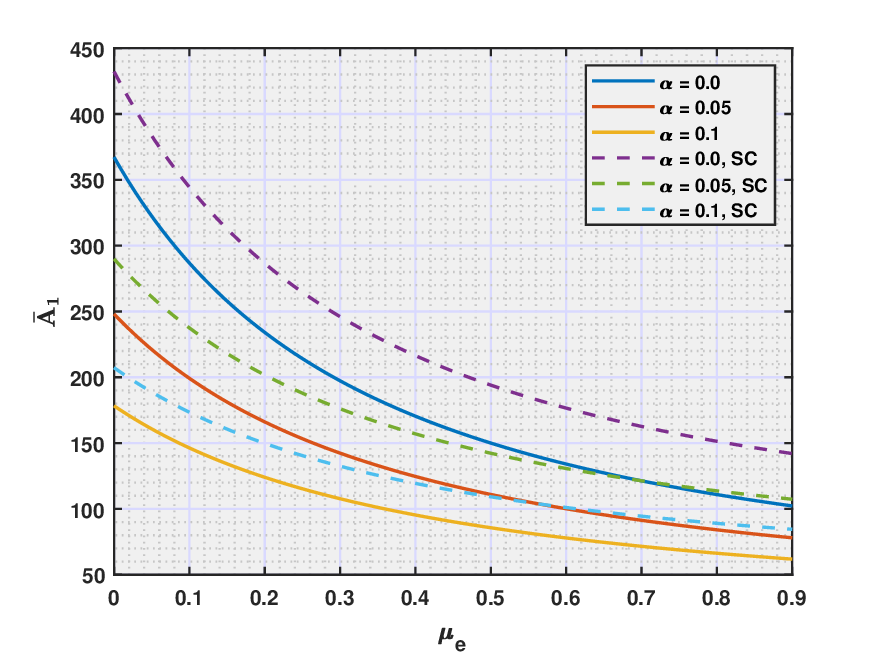}
        \caption{The variation of $\bar{A}_{1}$ with the non-thermal parameter ($\alpha$). Here we set, $n_{d0} \sim 10^{12}m^{-3}$, $Z_{d} \sim 10^{3}$ and $\gamma_{jd} = 1.5$}
        \label{A11vsMue1}
    \end{subfigure}
    \hfill
    \begin{subfigure}[b]{0.45\textwidth}
        \centering
        \includegraphics[width=\textwidth]{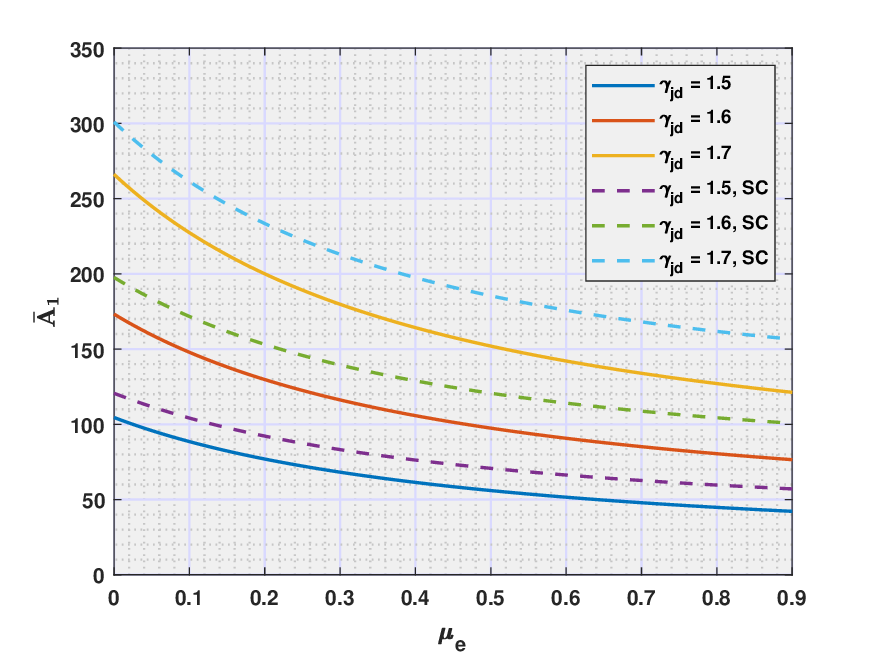}
        \caption{The variation of the $\bar{A}_{1}$ with the normalized Jeans frequency for dust ($\gamma_{jd}$). The other parameters being constant and have the values such as, $\alpha = 0.2$, $Z_{d} \sim 10^{3}$, and $n_{d0} \sim 10^{12}$.}
        \label{A11vsMue2}
    \end{subfigure}
    \caption{The figure shows the variation of nonlinear coefficient ($\bar{A}_{1}$) as a function of the electro-dust density ratio ($\mu_{e0}$). The nonlinear coefficient is required for the formation of solitons (which are stable, localised wave packets that retain their shape across extended distances). The nonlinearity causes the initial wave steepening, while the dispersive term keeps the wave from breaking, allowing it to grow into a soliton.}
    \label{A11vsMue}
\end{figure}

 In this problem, since we have derived a coupled KdV equation hence, both the KdV equations (i.e. equations (\ref{eqn34}) and (\ref{eqn35})) contains two pairs of nonlinear coefficients. One pair of nonlinear coefficients for the electrostatic potential which are symbolized as $A_{1}$ and $\bar{A}_{1}$ and the other pair corresponds to the gravitational potential which are denoted by $D_{1}$ and $\bar{D}_{1}$. Now, in section (\ref{solckdv}) we have derived the relations between the amplitudes of the two potentials and that of the velocity of solitons. It is interesting to note that the expressions of all these aforementioned parameters clears states that they have dependence mainly on $A_{1}$, $\bar{A}_{1}$, $B_{1}$ and $\bar{B}_{1}$. Hence, our next observation is to study the nature of variation of the nonlinear coefficients $A_{1}$ and $\bar{A}_{1}$ with respect to different plasma parameters. Thus we have plotted $A_{1}$ and $\bar{A}_{1}$ as a function of $\mu_{e}$ which are represented by the figures (\ref{A1vsMue}) and (\ref{A11vsMue}) respectively. Both the plots of figure (\ref{A1vsMue}) shows that the function $A_{1}$ undergoes a change of sign. With increase of the density ratio, $\mu_{e}$ the value of the nonlinear coefficient increases. Since, it starts from a negative value hence during its course of increment at certain value of $\mu_{e}$ it becomes zero. We denote this value of $\mu_{e}$ as the `cut-off' value. Now, the figure (\ref{A1vsMue3}) as well as the figure (\ref{A1vsMue4}) indicates that the `cut-off' value changes with the change of $\alpha$ or $\gamma_{jd}$ respectively. In figure (\ref{A1vsMue3}) we observe that both in absence or in presence of strong coupling the 'cut-off' value decreases as the value of $\alpha$ gets increases. This result indicates that the increase in number of the energetic ions causes the nonlinearity to vanish for lower value of electron density ($n_{e0}$). Similar nature is observed in figure (\ref{A1vsMue4}) when the value of $\gamma_{jd}$ got varied. But one interesting thing is observed in figure (\ref{A1vsMue4}) which clearly shows that in absence of strong coupling and for $\gamma_{jd} = 1.7$, $A_{1}$ is always remains positive. Hence, this typical value of $\gamma_{jd}$ plays a pivotal role in this case which shows that for this value of $\gamma_{jd}$ or for $\gamma_{jd} > 1.7$ the nonlinearity is not going to vanish and we will obtain a stable solitonic structure for the electrostatic potential. Similarly, we analyzed the nature of the nonlinear coefficient $\bar{A}_{1}$ by plotting it as a function of $\mu_{e}$, which are represented by the figure set (\ref{A11vsMue}). The figure (\ref{A11vsMue1}) shows that magnitude of $\bar{A}_{1}$ decreases with both $\mu_{e}$ and $\alpha$. But the value of $\bar{A}_{1}$ always remains positive. Infact, as $\mu_{e} \Rightarrow 0.9$, $\bar{A}_{1}$ tends to attain a constant value. Furthermore, when the variation of $\bar{A}_{1}$ with $\gamma_{jd}$ is observed (depicted by the figure (\ref{A11vsMue2})), we get a similar structure as obtained in the figure (\ref{A11vsMue1}). The only difference is, $\bar{A}_{1}$ increases with $\gamma_{jd}$. Over all, the figure set (\ref{A11vsMue}) shows that $\bar{A}_{1}$ always remains non-zero which supports that fact the the solitary structures arising due to the gravitational potential will never vanish whatever value the electron density or dust density attains. 
\par Actually, the nonlinear term describes how the wave speed depends on the amplitude of the wave. Specifically, in the KdV equation, larger amplitudes move faster than smaller ones, leading to a steepening effect. This nonlinear steepening means that as a wave propagates, its profile becomes increasingly asymmetric, with the steepening of the wavefront over time. The factor ``$A_1$ (or $\bar{A}_1$)" (the nonlinear coefficient) controls the strength of the nonlinearity in the system. Depending upon the value of this coefficient, the relationship between wave speed and amplitude would change, affecting the steepening process. In physical systems modeled by the KdV equation, the nonlinear coefficient also influences the energy transport properties of solitons. Solitons tend to carry energy and momentum, and the strength of the nonlinearity dictates how much energy is transported by each soliton. Hence, the variation of $A_1$ or  $\bar{A}_1$ with respect to $\mu_{e}$ are shown.

\par
\begin{figure}
    \centering
    \begin{subfigure}[b]{0.45\textwidth}
        \centering
        \includegraphics[width=\textwidth]{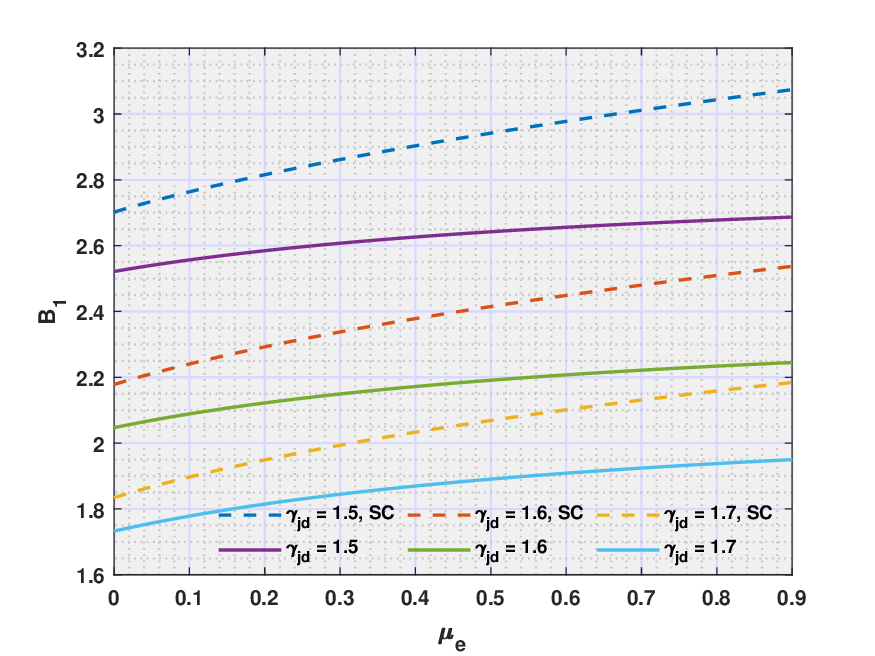}
        \caption{The variation of the coefficient $B_{1}$ with the non-thermal parameter ($\alpha$). Here we set, $n_{d0} \sim 10^{12}m^{-3}$, $Z_{d} \sim 10^{3}$ and $\gamma_{jd} = 1.5$}
        \label{B1vsMue}
    \end{subfigure}
    \hfill
    \begin{subfigure}[b]{0.45\textwidth}
        \centering
        \includegraphics[width=\textwidth]{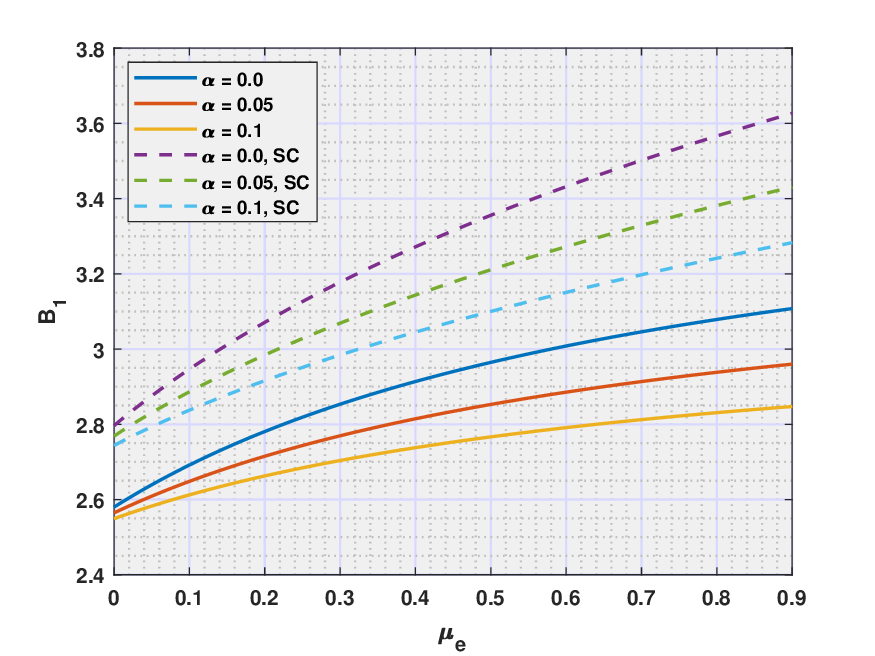}
        \caption{The variation of the coefficient $B_{1}$ with the normalized dust Jeans frequency ($\gamma_{jd}$). The other parameters being constant and have the values such as, $\alpha = 0.2$, $Z_{d} \sim 10^{3}$, and $n_{d0} \sim 10^{12}$.}
        \label{B1vsMue2}
    \end{subfigure}
    \caption{The plot of the time derivative coefficient $B_{1}$.}
    \label{B1_Mue}
\end{figure}

 If we focus on equation (\ref{eqn38}), we see that the amplitude of the soliton due to electrostatic potential, $a$ has a dependence on the coefficient $B_{1}$ as well as on the coefficient $\bar{B}_{1}$. Hence, it is necessary to see the nature of variation of these coefficients. Thus we have plotted $B_{1}$ versus $\mu_{e}$ shown by the figure set (\ref{B1_Mue}). We see that it in both figures i.e (\ref{B1vsMue}) and (\ref{B1vsMue2}) the value of $B_{1}$ always remains positive which is in accordance with the result (\ref{eqn37}) which states that $B_{1}$ cannot be negative since it has a proportional relation with the soliton velocity ($U$).  

\begin{figure}
    \centering
    \begin{subfigure}[b]{0.45\textwidth}
        \centering
        \includegraphics[width=\textwidth]{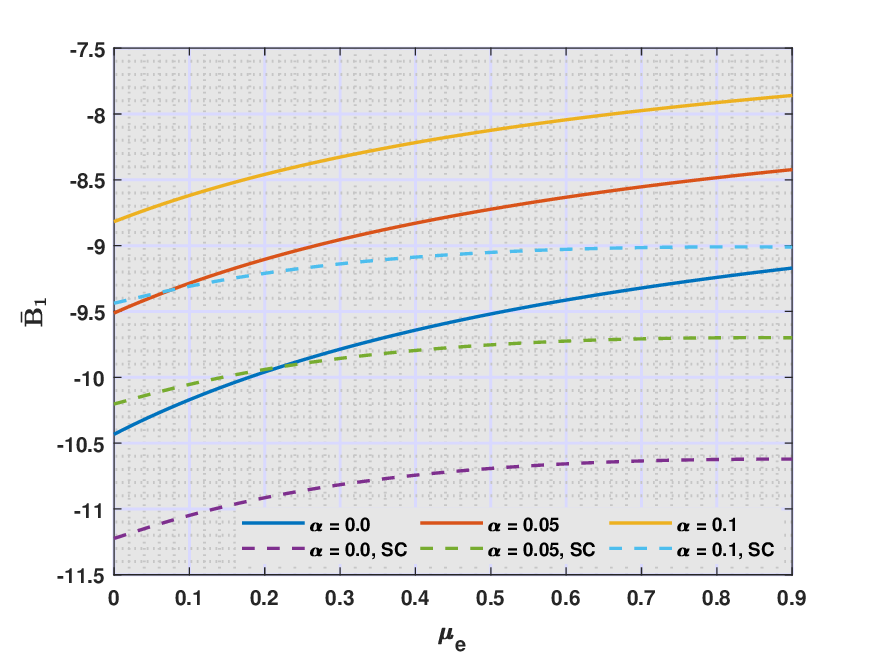}
        \caption{The variation of the coefficient $\bar{B}_{1}$ with the non-thermal parameter ($\alpha$). Here we set, $n_{d0} \sim 10^{12}m^{-3}$, $Z_{d} \sim 10^{3}$ and $\gamma_{jd} = 1.5$}
        \label{B11vsmue1}
    \end{subfigure}
    \hfill
    \begin{subfigure}[b]{0.45\textwidth}
        \centering
        \includegraphics[width=\textwidth]{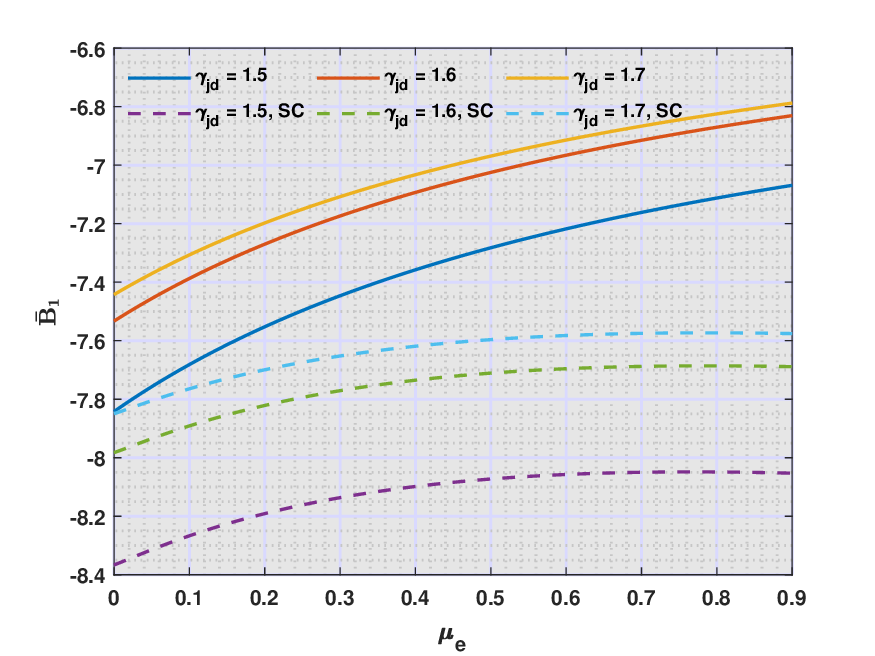}
        \caption{The variation of the coefficient $\bar{B}_{1}$ with the normalized dust Jeans frequency ($\gamma_{jd}$). The parameters which are kept constant, have the values such as, $\alpha = 0.2$, $Z_{d} \sim 10^{3}$, and $n_{d0} \sim 10^{12}$.}
        \label{B11vsmue2}
    \end{subfigure}
    \caption{The figure shows the variation of time derivative coefficient $\bar{B}_{1}$.}
    \label{B11vsmue}
\end{figure}

The equation (\ref{eqn40}) indicates that the amplitude of $\psi_{1}$-soliton i.e. $b$, is not only proportional to $a$ (the amplitude of $\phi_{1}$-soliton) but also to the coefficient $\bar{B}_{1}$. So, in order to ascertain the nature of variation of the amplitude of the solitary wave due to the gravitational potential we have plotted $\bar{B}_{1}$ with different plasma parameters represented by the figure set (\ref{B11vsmue}). The figure (\ref{B11vsmue1}) shows that with both $\mu_{e}$ and $\alpha$, $\bar{B}_{1}$ increases but the important part to note in this figure is the sign of $\bar{B}_{1}$ which is throughout negative both in absence or in presence of strong coupling. This indicates that the amplitude $a$ will be negative by virtue of equation (\ref{eqn38}). Though the magnitude and nature of variation are little bit different but $\bar{B}_{1}$ remains negative even when varied with $\gamma_{jd}$ which is depicted by the figure (\ref{B11vsmue2}). Hence, equations (\ref{eqn38}) and (\ref{eqn40}) demands that, the amplitude $a$ and $b$ both will also be negative. Thus, for both electrostatic and gravitational potential there will be a formation of dark soliton.

The time derivative term is critical in describing the evolution of the wave profile over time. The coefficient that multiplies the time derivative has important physical implications, particularly regarding wave speed and time scaling. The time derivative represents the rate of change of the wave profile at a given position over time. Its coefficient essentially dictates the timescale on which the evolution occurs. If the value of the coefficient changes, it would imply a rescaling of the time axis, meaning that the wave would evolve more slowly (if the value of the coefficient is greater than 1) or more quickly (if the value of the coefficient is lesser than 1).

\par
\begin{figure}
    \centering
    \begin{subfigure}[b]{0.45\textwidth}
        \centering
        \includegraphics[width=\textwidth]{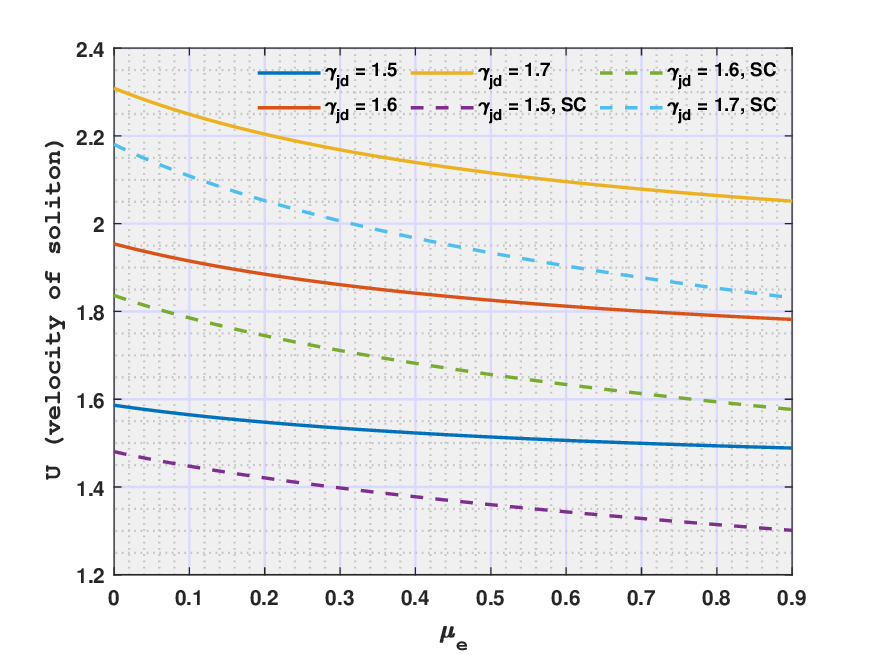}
        \caption{The variation of the phase velocity with the non-thermal parameter ($\alpha$). Here we set, $n_{d0} \sim 10^{12}m^{-3}$, $Z_{d} \sim 10^{3}$ and $\gamma_{jd} = 1.5$}
        \label{v_vs_Mue}
    \end{subfigure}
    \hfill
    \begin{subfigure}[b]{0.45\textwidth}
        \centering
        \includegraphics[width=\textwidth]{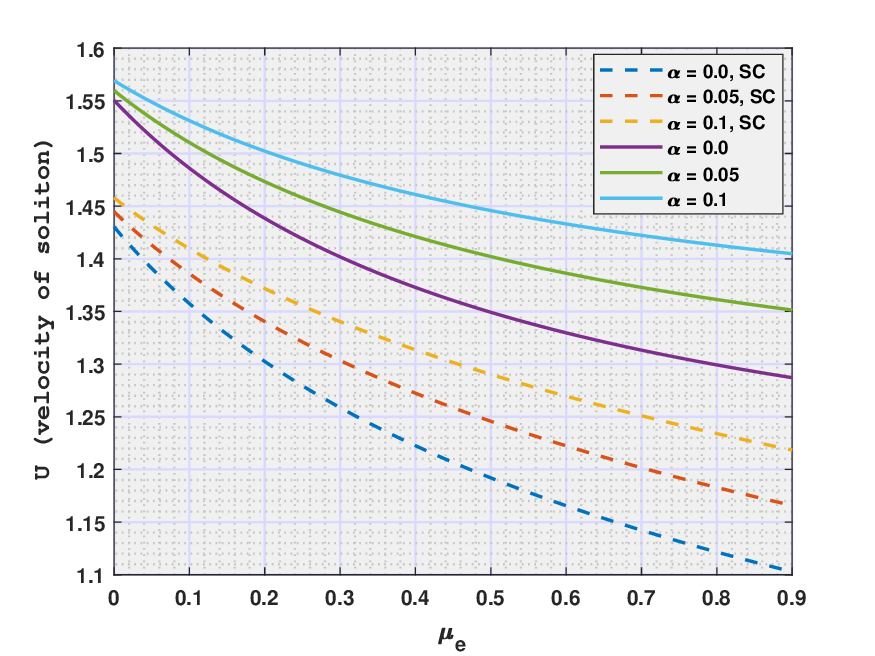}
        \caption{The variation of the phase velocity with the normalized dust Jeans frequency ($\gamma_{jd}$). The other parameters being constant and have the values such as, $\alpha = 0.2$, $Z_{d} \sim 10^{3}$, and $n_{d0} \sim 10^{12}$.}
        \label{v_vs_Mue1}
    \end{subfigure}
    \caption{The figure shows the variation of phase velocity ($v_{p}$) as a function of the dust charge number ($Z_{d}$).}
    \label{solitonvelocity}
\end{figure}

 Since, the wave amplitudes have a dependence on the velocity of the soliton ($U$) thus it is necessary to plot the variation of the $U$ for different plasma parameters. Thus figure set (\ref{solitonvelocity}) represents the variational nature of the soltonic velocity ($U$) for both presence and absence of strong coupling and for various plasma parameters. Both the figures (\ref{v_vs_Mue}) and (\ref{v_vs_Mue1}) indicate that the velocity of soliton decreases with the electron number density ($n_{e0}$). The figure (\ref{v_vs_Mue}) depicts that with increase of $\gamma_{jd}$ the magnitude of soliton velocity increases but eventually as $\mu_{e} \rightarrow 0.9$, $U$ becomes constant. Similar situation is obtained when the variation is observed when $U$ is varied with $\alpha$, depicted by figure (\ref{v_vs_Mue1}).

\begin{figure*}
    \centering
    \includegraphics[width=1.0\textwidth,]{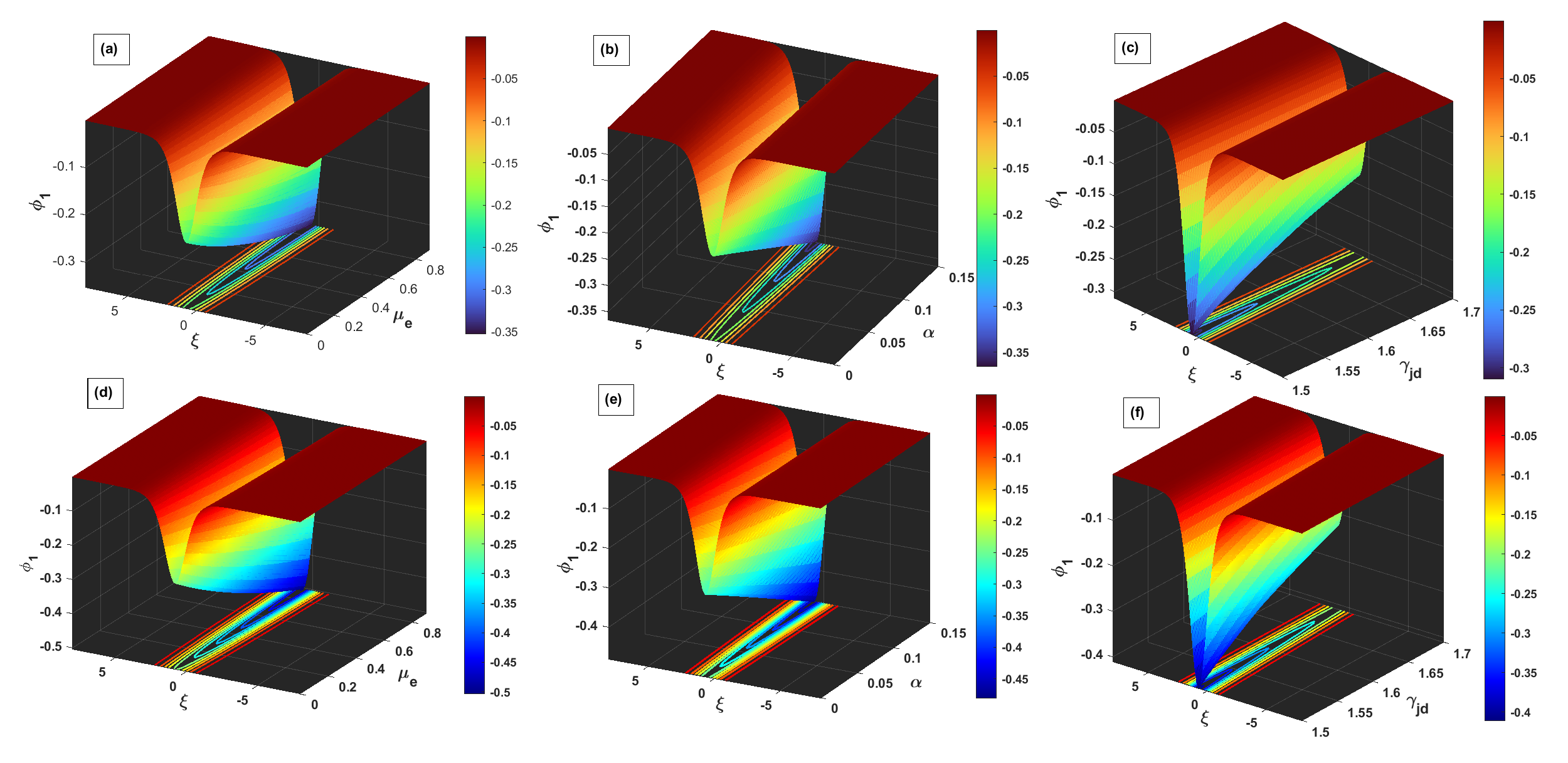}
    \caption{A 3D plot of the dark soliton due to the electrostatic potential. The figure shows the variation of the solitary wave profile with $\mu_{e}$, $\alpha$ and $\gamma_{jd}$ both in absence and presence of strong coupling.}
    \label{darksolitonsubplot1}
\end{figure*}

\begin{figure*}
    \centering
    \includegraphics[width=1.0\textwidth,]{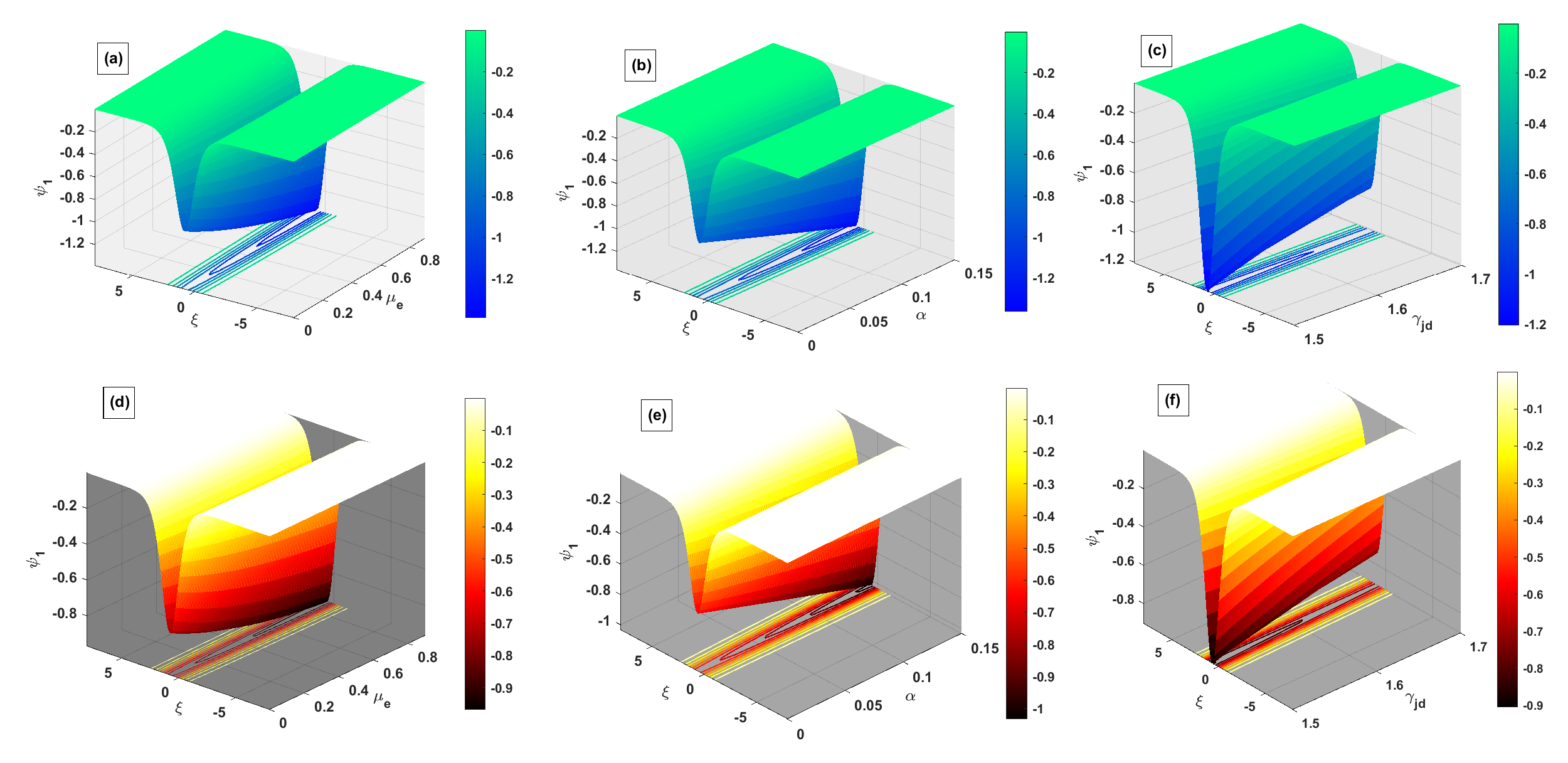}
    \caption{A 3D plot of the dark soliton due to the gravitational potential. The figure shows the variation of the solitary wave profile with $\mu_{e}$, $\alpha$ and $\gamma_{jd}$ both in absence and presence of strong coupling.}
    \label{darksolitonsubplot2}
\end{figure*}

Finally, to see how the effect of strong coupling, the nonthermality of ions, the gravitational field etc. effect the soliton propagation, we plotted the solitary wave profiles as a function of different plasma parameters which is presented by the figure sets (\ref{darksolitonsubplot1}) and (\ref{darksolitonsubplot2}). The figure (\ref{darksolitonsubplot1}) refers to the solitary wave profile originating due to electrostatic potential. The first row of the figure (\ref{darksolitonsubplot1}) i.e (a, b \& c) denotes the propagation of the solitary wave for the case when there is no strong coupling effect between the dust particles. While the figures (\ref{darksolitonsubplot1} e, f \& g) represents the propagation characteristics of the wave when the dust particles are considered to be strongly coupled. The figure (\ref{darksolitonsubplot1} a), represents how the propagation of dark soliton is affected by the electron number density. Since, the dust number density constant, hence increase in value of $\mu_{e}$ implies increase in the value of electron number density. Now, the increase in depth of the dark soliton corresponds to stronger phase change across the soliton. Interestingly, we see in the figures (\ref{darksolitonsubplot1} a, b, d, e) the depth increases with both $\mu_{e}$ and $\alpha$. Thus, increase in the number density of the electrons and increase in the number of energetic ions enhances the phase change of soliton for both presence and absence of strong coupling. On the other-hand the figures (\ref{darksolitonsubplot1} c \& f) indicates that with the increase of the Jeans frequency of the dust the depth of the dark soliton decreases which implies that the soliton tends to become more stability. The subplots of the upper and lower row of the figure (\ref{darksolitonsubplot1}) differs only in the magnitudinal aspect. Thus we can conclude that the effect of strong coupling in this case just alters the numerical value of the variables related to the nonlinear wave excitations.  

Last but not the least, the figure set (\ref{darksolitonsubplot2}) shows the solitary wave profile for the dark soliton formed due to the gravitational potential ($\psi_{1}$). Though the nature of variation is same as that of $\phi$-dark solitons but the differences are in the magnitude of depth of the dark soliton formed in this case. This is obvious because of equation (\ref{eqn40}). It is quite evident that dark soliton with different depth will have different intensities. Generally, dark solitons with greater depth travel faster than that with lesser depth. This due to that fact that deeper  soliton have more more energies which make them possible to propagate more quickly. Hence, in this case by observing the figure sets (\ref{darksolitonsubplot1}) and (\ref{darksolitonsubplot2}) we infer that the $\psi_{1}$ solitons will travel more faster than the $\phi_{1}$ solitons.

\section{Conclusion}
In the above paper we have tried to analyse the mutual influence of the strong coupling and gravitational interaction in a three component dusty plasma. Even in the linear domain we have seen that the phase velocity strongly depends on the presence or absence of strong coupling and the number of ions deviating from the thermal equilibrium. It increases with the Jeans frequency. But when we pass over to the nonlinear domain via reductive perturbation we observe that a coupled set of KdV equation arises and the most interesting part to be noted is that we always encounter dark soliton. In the statndard strongly coupled plasma such is always the case. It is observed that the dark soliton corresponding to the gravitational potential is always larger in magnitude. Infact self-gravitating, strongly coupled dusty plasmas are important for understanding processes such as planetary formation, cosmic dust dynamics, and even phenomena in space weathering. The complex interaction of gravitation, electrostatic forces, and plasma dynamics makes these systems rich subjects for both theoretical and experimental research.

\appendix
\begin{equation}\label{E0}
  \theta_1 = -\sigma_i + \Lambda_i
\end{equation}
\begin{equation}\label{E1}
  \theta_2 = (\sigma_i^2)/ - \sigma_i + \Lambda_i^2
\end{equation}
\begin{equation}\label{a0}
  F = (v_p^2) - (d_0 + d_{11})
\end{equation}
\begin{equation}\label{a1}
\begin{split}
  &A_1 = 2(-1/F)(d_{23}((\gamma_{ji}/\gamma_{jd})^4)(\theta_1^2) + d_{24}((\gamma_{ji}/\gamma_{jd})^2)2\theta_1^2 \\
  &- d_{25}^2 + ((\gamma_{ji}/\gamma_{jd})^4)(\theta_1^2) - d_{12}*((\gamma_{ji}/\gamma_{jd})^2)\theta_1 + \\ &d_0((\gamma_{ji}/\gamma_{jd})^4)*(\theta_1^2) - d_{11}((\gamma_{ji}/\gamma_{jd})^2)(\theta_1^2)^2 + \\ &d_{12}((\gamma_{ji}/\gamma_{jd})^2)\theta_1^2 + (v_p^2)((\gamma_{ji}/\gamma_{jd})^2)\theta_1^2(d_{12}-1)) \\
  &- \sigma_e^2 - 2\mu_i\theta_2
\end{split}
\end{equation}
\begin{equation}\label{a2}
  B_1 = (1/F)(v_p((\gamma_{ji}/\gamma_{jd})^2)\theta_1 - (v_p/F)(d{12} -1))
\end{equation}
\begin{equation}\label{a3}
\begin{split}
  C_1 = ((v_p^{2})/(F^2))(((\gamma_{ji}/\gamma_{jd})^2)\theta_1 - \\
  (d_{12}-1)/F + (v_p/F)(d_{12}-1)^2)
  \end{split}
\end{equation}
\begin{equation}\label{a4}
  D_1 = ((v_p^{2})/(F^3))
\end{equation}
\begin{equation}\label{a5}
  G_1 = v_p/F
\end{equation}

\begin{equation}\label{a6}
  \bar{G}_{1}= -\gamma_{jd}^2*v_p/(v_p^2-(d_0+d_{11}))^2
\end{equation}

\begin{equation}\label{a7}
  \bar{B}_{1} = (\gamma_{ji}/\gamma_{jd})^2\theta_1 - \bar{G}_{1}(d_{12}-1)
\end{equation}

\begin{equation}\label{a8}
\begin{split}
  \bar{C}_{1} = -\bar{G}_{1}v_p\theta_1(\gamma_{ji}/\gamma_{jd})^2 +\\
  v_p(d_{12}-1)\bar{G}_{1}/(v_p^2-(d_0+d_{11}))
  \end{split}
\end{equation}

\begin{equation}\label{a9}
  \bar{D}_{1} = \bar{G}_{1}v_p/(v_p^2-(d_0+d_{11}))
\end{equation}

\begin{equation}\label{a10}
\begin{split}
  \bar{A}_{1} = (\bar{B}_{1}A_1 - 2(\bar{B}_{1}/B_1)(C_1 \bar{B}_{1} - \bar{C}_{1}B_1) - \\
  ((\bar{B}_{1}/B_1)^2)(C_1 \bar{B}_{1} - \bar{C}_{1}B_1))/B_1;
  \end{split}
\end{equation}

\section*{Acknowledgements}
One of the author Dr. Shatadru Chaudhuri (SC) would like to express his sincere gratitude to his supervisor Dr. A Roy Chowdhury, for his invaluable guidance and support throughout this research. He would also like to thank his colleague Dr. Shahin Nasrin for her insightful discussions and collaboration. Special thanks to Dr. Papiya Chaudhury, principal of Shri Shikshayayan college, for providing a conducive work environment that fosters research and academic growth. Her support and resources have been instrumental in enabling the pursuit of knowledge and innovation.

\section*{Declaration}
\begin{enumerate}
\item All authors contributed equally.
\item There is no conflict of interest.
\item This research study received no external funding.
\item The manuscript has no associated data since the expressions are evaluated using MATLAB software.
\end{enumerate}


\section{Biography Section}
 

\begin{IEEEbiography}[{\includegraphics[width=1in,height=1.25in,clip,keepaspectratio]{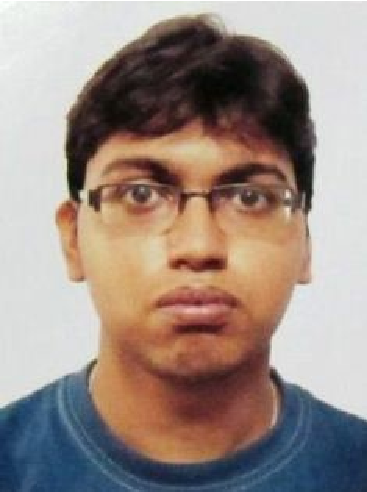}}]{Shatadru Chaudhuri} recieved the Bachelor of Science degree from Calcutta University and Master of Science degree in Physics from Jadavpur University. He completed and got awarded his Ph.D degree in Plasma Physics from Department of Physics of Jadavpur University. Now, he is a faculty in the Department of Physics in Shri Shikshayatan College, Kolkata, India.
\end{IEEEbiography}

\begin{IEEEbiography}[{\includegraphics[width=1in,height=1.25in,clip,keepaspectratio]{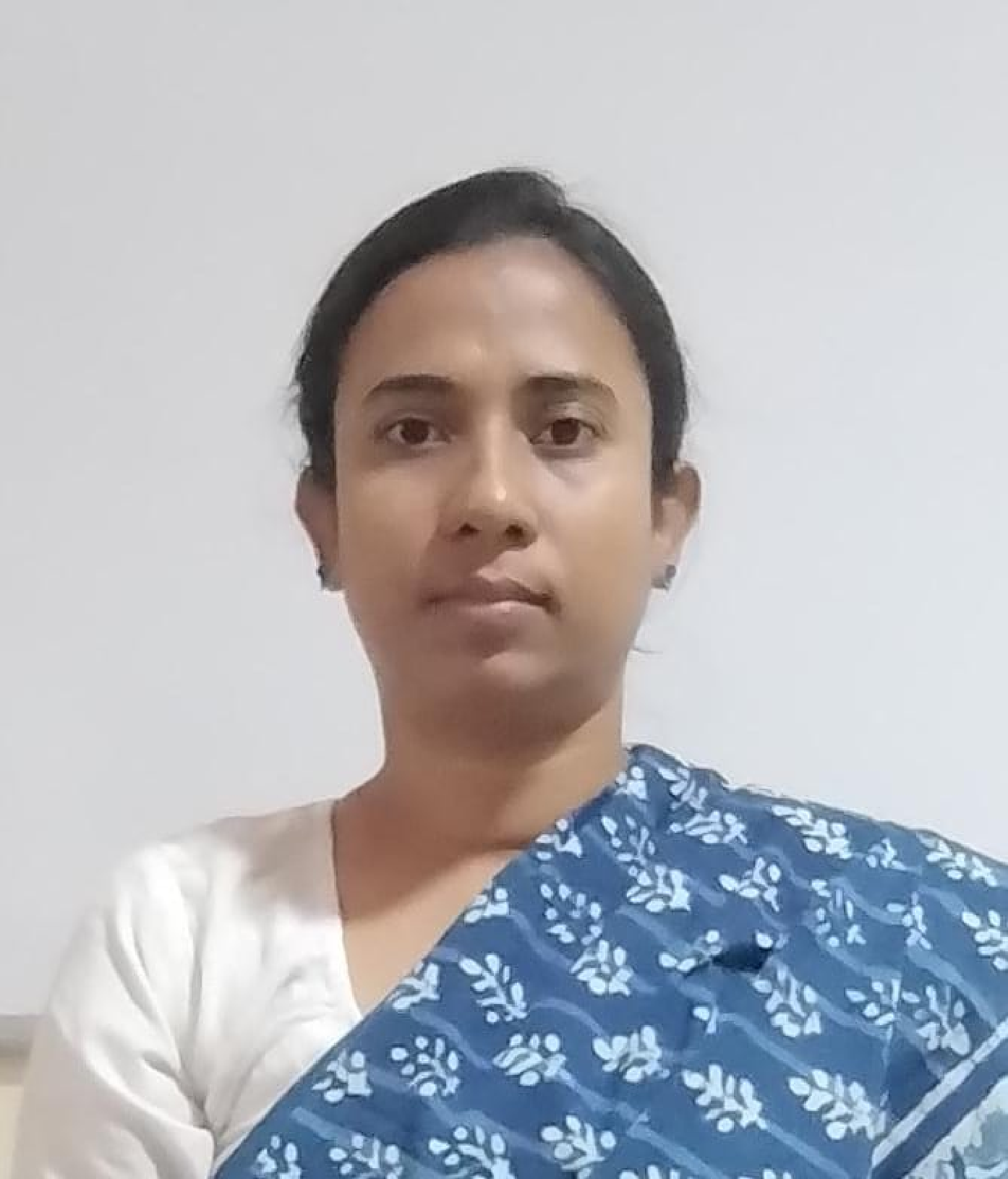}}]{Shahin Nasrin} received the B.Sc. degree in Physics from Calcutta University, Kolkata, India, in 2008. She acheived the M.Sc. degree in applied Physics from BESU, Shibpur, India in 2010 and the Ph.D degree froem Jadavpur University, Kolkata for her work on cross-field interaction in Plasma in the year 2023. Currently she is a senior faculty and the Head of the Department of Shri Shikshayatan College, Kolkata.

\end{IEEEbiography}

\begin{IEEEbiography}[{\includegraphics[width=1in,height=1.25in,clip,keepaspectratio]{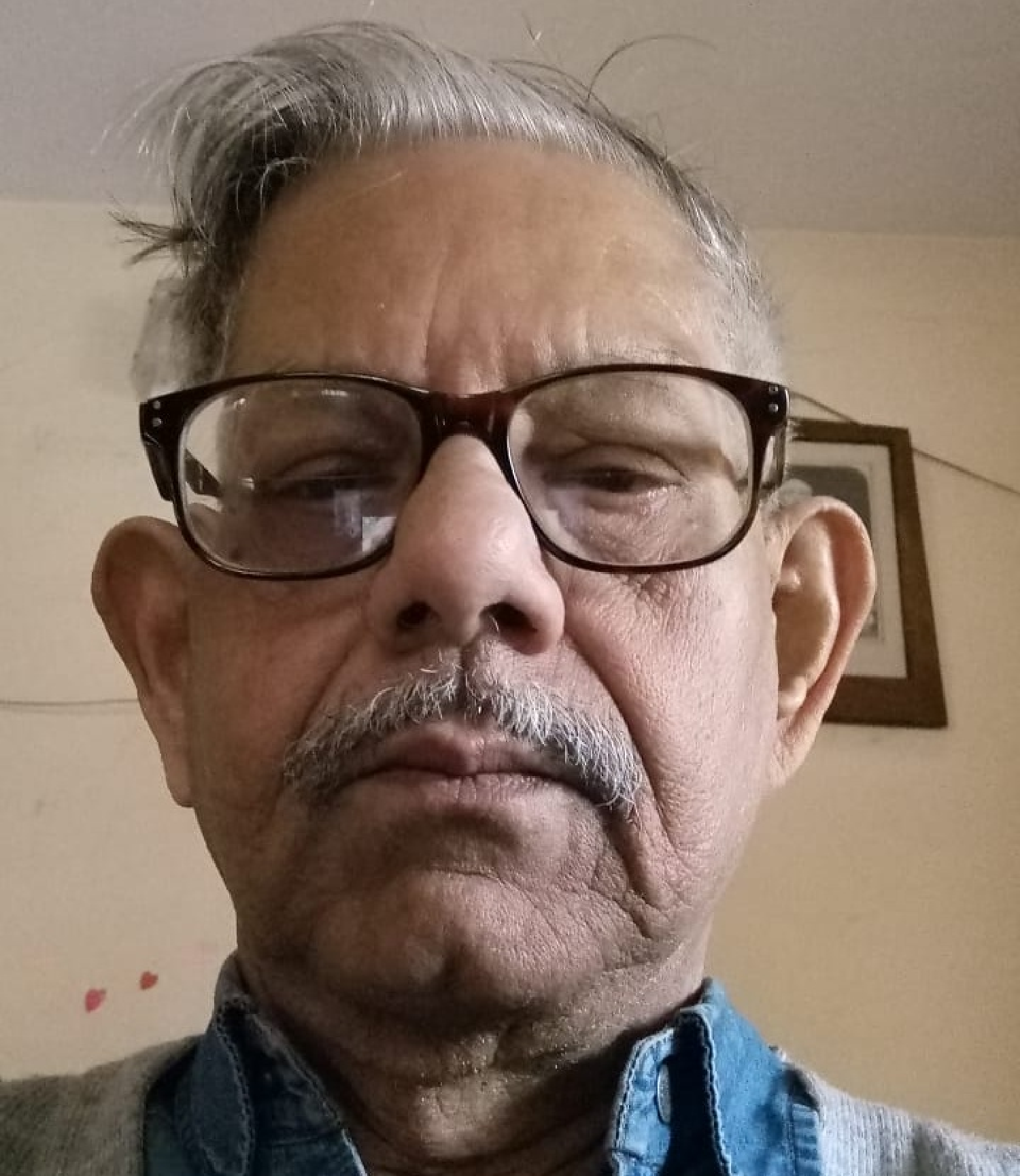}}]{A Roy Chowdhury}
 is now Professor Emeritus at the department of Physics, Jadavpur University . He was an associate member of ICTP, Triest from 1982 to 1988. He has done extensive research work on Plasma Physics, Nonlinear waves, Soliton, Integrable system, Nonlinear Optics and Chaos. He graduated from Presidency College, Calcutta University with Physics and Mathematics in the year 1964. He did his Ph.D on the High Energy Physics from Jadavpur University, Kolkata. He was European Economic Community Fellow to Paderborn, Germany in the year 1989. He has published four books from Chapman and Hall, U.K.
\end{IEEEbiography}

\clearpage
\end{document}